\documentclass[aps,prl,12pt,onecolumn,showpacs,superscriptaddress,floatfix]{revtex4-2}
\usepackage{amssymb, amsmath}
\usepackage{graphicx}
\usepackage{xfrac, stmaryrd, trimclip}
\usepackage{bm}
\usepackage{gensymb}
\usepackage[normalem]{ulem}
\usepackage{xcolor}
\usepackage[colorlinks,linkcolor=blue,anchorcolor=blue,citecolor=blue,urlcolor=blue,filecolor=blue,menucolor=blue,runcolor=blue]{hyperref}%

\newcommand{\KV}{KV$_2$Se$_2$O}
\newcommand{\SM}{SmB$_6$}

\begin{document}
\title{Visualizing spin-polarization of an altermagnet \KV\ via spin-selective tunneling}

\author{Guofei Yang}
\thanks{These authors contributed equally to this work.}
\affiliation{New Cornerstone Science Laboratory, Center for Correlated Matter and School of Physics, Zhejiang University, Hangzhou 310058, China}

\author{Chuang Li}
\thanks{These authors contributed equally to this work.}
\affiliation{New Cornerstone Science Laboratory, Center for Correlated Matter and School of Physics, Zhejiang University, Hangzhou 310058, China}

\author{Chengwei Wang}
\thanks{These authors contributed equally to this work.}
\affiliation{New Cornerstone Science Laboratory, Center for Correlated Matter and School of Physics, Zhejiang University, Hangzhou 310058, China}

\author{Xudong Zhao}
\affiliation{New Cornerstone Science Laboratory, Center for Correlated Matter and School of Physics, Zhejiang University, Hangzhou 310058, China}

\author{Yifan Wan}
\affiliation{New Cornerstone Science Laboratory, Center for Correlated Matter and School of Physics, Zhejiang University, Hangzhou 310058, China}

\author{Hengrui Gui}
\affiliation{New Cornerstone Science Laboratory, Center for Correlated Matter and School of Physics, Zhejiang University, Hangzhou 310058, China}

\author{Guoqing Zeng}
\affiliation{New Cornerstone Science Laboratory, Center for Correlated Matter and School of Physics, Zhejiang University, Hangzhou 310058, China}

\author{Saizheng Cao}
\affiliation{New Cornerstone Science Laboratory, Center for Correlated Matter and School of Physics, Zhejiang University, Hangzhou 310058, China}

\author{Chuqiao Hu}
\affiliation{State Key Laboratory of Optoelectronic Materials and Technologies,
Guangdong Province Key Laboratory of Display Material and Technology, and
School of Electronics and Information Technology, Sun Yat-sen University, Guangzhou 510275, China.}

\author{Qihe Yu}
\affiliation{Hubei Key Laboratory of Photoelectric Materials and Devices, School of Materials Science and Engineering,
Hubei Normal University, Huangshi 435002, China}

\author{Yujia Zhang}
\affiliation{Hubei Key Laboratory of Photoelectric Materials and Devices, School of Materials Science and Engineering,
Hubei Normal University, Huangshi 435002, China}

\author{Dong Chen}
\affiliation{College of Physics, Qingdao University, Qingdao 266071, China}

\author{Yu Liu}
\affiliation{New Cornerstone Science Laboratory, Center for Correlated Matter and School of Physics, Zhejiang University, Hangzhou 310058, China}

\author{Yu Song}
\affiliation{New Cornerstone Science Laboratory, Center for Correlated Matter and School of Physics, Zhejiang University, Hangzhou 310058, China}

\author{Yongjun Zhang}
\affiliation{Hubei Key Laboratory of Photoelectric Materials and Devices, School of Materials Science and Engineering,
Hubei Normal University, Huangshi 435002, China}

\author{Fei Liu}
\email[Corresponding author: ]{liufei@mail.sysu.edu.cn}
\affiliation{State Key Laboratory of Optoelectronic Materials and Technologies,
Guangdong Province Key Laboratory of Display Material and Technology, and
School of Electronics and Information Technology, Sun Yat-sen University, Guangzhou 510275, China.}

\author{Lun-Hui Hu}
\email[Corresponding author: ]{lunhui@zju.edu.cn}
\affiliation{New Cornerstone Science Laboratory, Center for Correlated Matter and School of Physics, Zhejiang University, Hangzhou 310058, China}

\author{Lin Jiao}
\email[Corresponding author: ]{lin.jiao@zju.edu.cn}
\affiliation{New Cornerstone Science Laboratory, Center for Correlated Matter and School of Physics, Zhejiang University, Hangzhou 310058, China}

\author{Huiqiu Yuan}
\email[Corresponding author: ]{hqyuan@zju.edu.cn}
\affiliation{New Cornerstone Science Laboratory, Center for Correlated Matter and School of Physics, Zhejiang University, Hangzhou 310058, China}
\affiliation {Institute of Fundamental and Transdisciplinary Research, Zhejiang University, Hangzhou 310058, China}
\affiliation  {State Key Laboratory of Silicon and Advanced Semiconductor Materials, Zhejiang University, Hangzhou 310058, China}

\begin{abstract}
\textbf{Altermagnetism, a recently identified magnetic phase that combines vanishing net magnetization with momentum-dependent spin splitting, challenges the conventional dichotomy between ferromagnets and antiferromagnets. While several candidate materials have been proposed,  microscopic  experimental evidence linking crystal symmetry, electronic structure and d-wave spin polarization remains scarce. Here we report the visualization of a metallic d-wave altermagnet in \KV. Through spin-selective scanning tunneling microscopy powered by a topological insulator tip,  we resolve temperature-dependent phase shift of Friedel oscillations and bias-dependent quasiparticle-interference between $q_x$ and $q_y$ directions. The W-tip controls instead show mostly in-phase oscillations and a non-sign-reversing QPI anisotropy. Together with numerical simulations, these observations support momentum-dependent spin polarization that follows a characteristic $d$-wave form factor. Our results establish \KV\ as a  material platform to study the interplay between spin-valley locking, Fermi-surface instability and unconventional magnetism, and highlight spin-selective tunneling as a local probe of compensated magnetic spin textures.}
\end{abstract}

\maketitle

The discovery of non-relativistic spin-split collinear altermagnets~\cite{vsmejkal2020crystal,Naka19NC,hayami2019momentum,ifmmode2022Beyond,jungwirth2025altermagnetism,song2025altermagnets,jungwirth2026symmetry} has marked a turning point in the study of unconventional magnetism~\cite{Hirsch1990prb,Ikeda1998prl,wu2004prl,CJWu07PRB,Chen2014Anomalous,liu2025different,liu2025prx,liu2025symmetry}. 
Unlike conventional N\'eel antiferromagnetism (AFM), altermagnets (AM) combine compensated magnetic sublattices with momentum-dependent spin splitting~\cite{ifmmode2022Emerging,bai2024altermagnetism,fender2025altermagnetism}, offering a pathway to electrically generate and manipulate spin currents without stray fields or net magnetization~\cite{Naka19NC,gonzal2021ruo2prl,bai_observation_2022,bose_tilted_2022,karube_observation_2022}. This new paradigm has rapidly advanced from theoretical prediction to experimental realization in materials such as RuO$_2$~\cite{fedchenko2024ruo2}, MnTe~\cite{amin_nanoscale_2024,krempasky2024altermagnetic,lee2024MnTe}, and CrSb~\cite{reimers2024CrSb,zhou2025manipulation,yang2025three}. Against this broader progress, the van der Waals $A$V$_2$Se$_2$O and $A$V$_2$Te$_2$O family ($A$ = Alkali metals)~\cite{jiang2025metallic,zhang2025crystal,hu2025pronounced,chen2025compression,Liu2025prbPhysical,Sun2025antiferrom,yang2025observation,hu2026observation} have emerged as a highly promising material platform due to their favorable metallic properties and sample dimensions, identified as a $d$-wave altermagnet in its monolayer~\cite{ma2021multifunctional} and a hidden altermagnet in the bulk~\cite{yang2025observation}. Within this material family, the Lieb lattice geometry of the V$_2$O plane gives rise to spin-polarized Fermi surfaces with quasi-1D flat segments, manifesting as a distinctive electronic structure that enables nearly 100\% efficiency in charge-to-spin current conversion~\cite{Lai2025dwave,li2026JJevenodd}.

The band structure of KV$_2$Se$_2$O has been investigated by angle-resolved photoemission spectroscopy (ARPES)~\cite{jiang2025metallic}. Yet, spin-resolved ARPES measurements are limited by magnetic domains and the inherently low efficiency of spin detection~\cite{zhang2022Angleresolved,liuSymmetry2026}. Until now,  microscopic evidence for AM-related spin splitting is still highly desirable. Complementary real-space techniques capable of visualizing momentum-dependent spin polarization at atomic resolution remain challenging. Moreover, altermagnetism inherently evades conventional magnetic probes: its compensated order yields no net magnetization, while standard spin-polarized tunneling methods via ferromagnetic tips can generate stray fields that locally perturb the delicate spin arrangement, potentially affecting the measurement~\cite{bodeSpin2003,wiesendangerSpinMappingNanoscale2009}. Therefore, a probe that minimizes such perturbations while maintaining the sensitivity to spin polarization is essential for reliably resolving the intrinsic spin structure.

Topological Kondo insulators offer a promising solution to this challenge. In SmB$_6$, strong correlations drive band inversion and produce Dirac surface states with helical spin–momentum locking~\cite{Dzero2010Topo}. When fabricated as a nanowire scanning tunneling microscopy (STM) tip, SmB$_6$ acts as an intrinsic directional spin filter~\cite{aishwarya2022spin,banerjee_axionic_2025}: the tunneling current is spin-polarized and the spin orientation could be flipped by bias voltage, rather than by an external magnetic  field~\cite{wiesendangerSpinMappingNanoscale2009}. 
In this work, we use SmB$_6$ nanowire–based STM probes to investigate the surface electronic structure of the AM material \KV. Using this topological tunneling geometry, we observe real-space Friedel oscillations and bias-dependent momentum-space QPI signatures that are consistent with symmetry-protected AM spin polarization. Together with W-tip controls experiments and theoretical simulations, these measurements provide microscopic evidence for spin-selective tunneling from the AM-derived electronic states.

\begin{figure}[h]
\includegraphics[width=0.8\columnwidth]{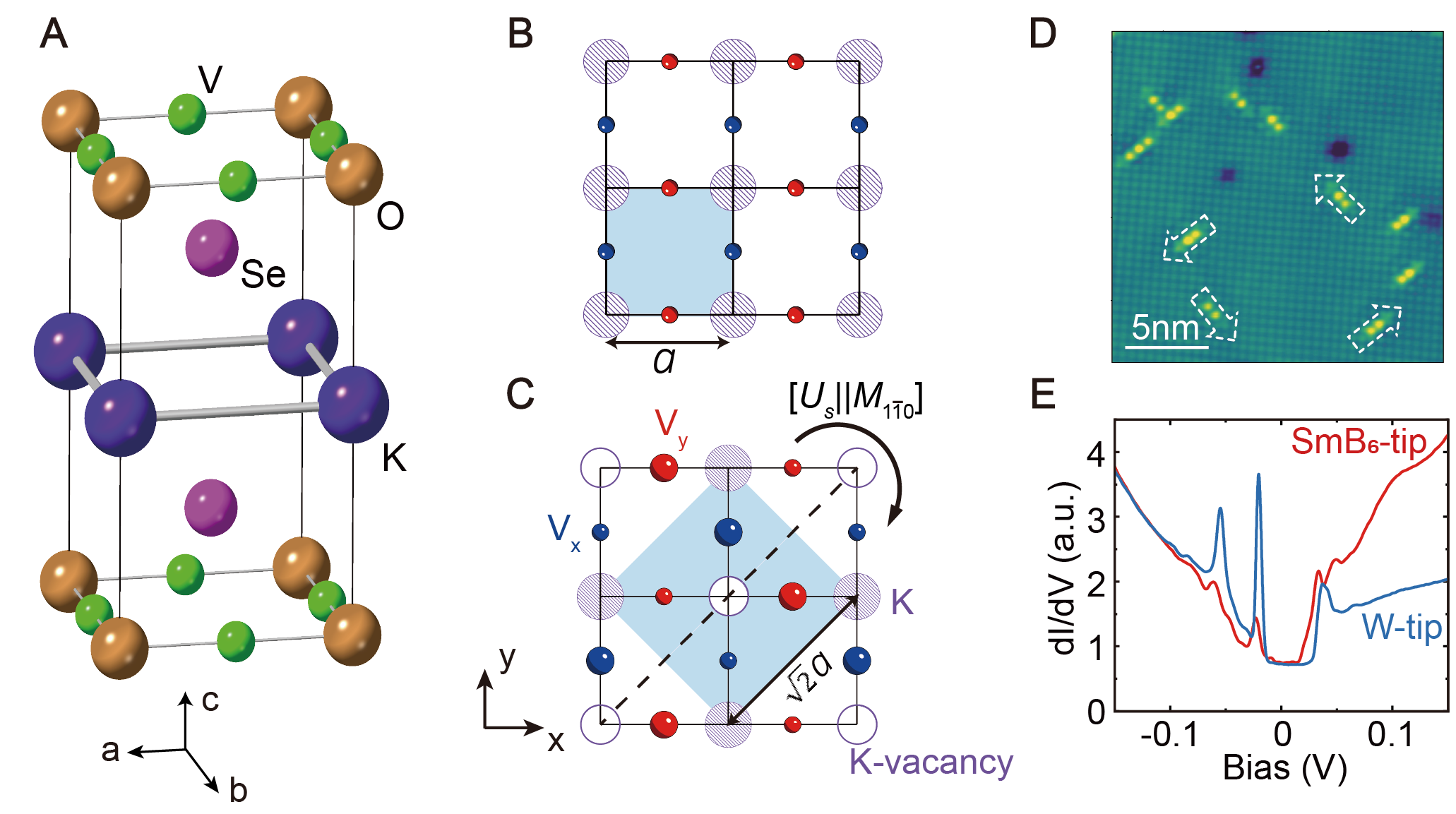}

\caption{\textbf{Crystal structure and basic surface properties.} 
{\bf (A)} Unit cell of \KV. 
{\bf (B)} Projection of K and V-plane with the shade purple circles represent K atoms, blue and red balls represent V atoms with local spin point down and up. Lattice constant $a$ = 3.966~\AA.
{\bf (C)} The cleaved surface of \KV\ with  half of the K atoms missing (empty circles). The residual K atoms form a reconstructed $\sqrt{2}a \times \sqrt{2}a$ square lattice. The different size of the red and blue balls represent an SDW modulation. Two opposite-spin sublattices are connected by a spin group symmetry $[U_s||M_{1\bar{1}0}]$. 
{\bf (D)} STM topography of the reconstructed  $\sqrt{2}a \times \sqrt{2}a$ K-terminated surface.  Dominant paired-protrusion defects exhibit four approximately orthogonal arrow-like orientations. 
{\bf (E)} d$I$/d$V$-spectrum obtained on the K-terminated surface, red and blue curves are measured at 5~K by W- and SmB$_6$-nanowire tips, respectively. 
}
\label{fig1}
\end{figure}

\KV\ was synthesized using KSe as the self-flux agent. The crystal structure is shown in Fig.~\ref{fig1}A, which has a tetragonal structure with a space group $P4/mmm$. The structure consists of V$_2$O, Se, and K layers stacked sequentially along the $c$-axis.
Specifically, the V$_2$O plane realizes a Lieb lattice geometry, in which the two symmetry-inequivalent V atoms within each unit cell are related by a four-fold rotation about the $c$-axis ($C_{4z}$) or by a mirror reflection through the $[1\bar{1}0]$ plane ($M_{1\bar{1}0}$). In the AM phase, these two V sites carry opposite local moments, and the magnetic structure remains invariant under the spin-space symmetries $[U_s||C_{4z}]$ and $[U_s||M_{1\bar{1}0}]$, where $U_s$ denotes a global spin flip. These symmetries impose the following constraints on the band structure: $E_{\uparrow}(k_x,k_y) = E_{\downarrow}(-k_y, k_x)$ and $E_{\uparrow}(k_x,k_y) = E_{\downarrow}(k_y, k_x)$. Spin degeneracy is therefore lifted at arbitrary momentum and is enforced only along the symmetry-protected diagonal lines $k_x=\pm k_y$. This symmetry characterization identifies the monolayer KV$_2$Se$_2$O as a $d$-wave AM.

Figure~\ref{fig1}B provides a top-down view (projection) of the K and V$_2$O planes. 
The local magnetic moments on each V site are oriented along the $c$-axis and are distinguished by red and blue colors, indicating an in-plane antiferromagnetic order, as established by neutron diffraction measurements~\cite{Sun2025antiferrom,yang2025observation}.
Sample cleaves in the K plane and leaves almost half of the K atoms on the terminated surface. The residual K atoms undergo surface reconstruction, forming a $\sqrt{2}a \times \sqrt{2}a$ lattice, as presented in Fig.~\ref{fig1}C.
Below N\'eel temperature $T_N$, a second phase transition occurs around $T_N'=96$ K (see Fig.  S12). 
Spin susceptibility measurements indicate the onset of a spin-density wave (SDW) order, although this is still under debate~\cite{jiang2025metallic,Sun2025antiferrom,yan2026magnetic,baiAbsence2024,zhuangCharge2025a}. The spin-polarized Fermi surfaces consist of nearly quasi-1D flat segments located around $k_x=\pm \pi/2a $ and $k_y=\pm \pi/2a$. A nesting vector $\bm{Q}=(\pi/a,\pi/a)$—favorable for an instability in the electron–hole channel—should connect Fermi surfaces of the same spin polarization to drive the formation of an SDW gap~\cite{xu2025prb}. Further details and modeling are provided in the Supplementary Materials (SM). Consequently, below $T_N'$, the electronic structure of the V$_2$O plane also reconstructs into a $\sqrt{2}a \times \sqrt{2}a$ superlattice.
Therefore, at low temperature, the observed topography of the K-lattice and the underlying spin structure of V$_2$O-magnetic lattice exhibit the same periodicity. Consequently, the bare checkerboard periodicity is not independently resolved in the K-terminated STM topography.

To experimentally probe the AM state in real space, we employ spin-sensitive STM to detect quasi-particle scattering off individual impurities. In this approach, any observed spin polarization in the scattered signal must arise intrinsically from the AM host. We therefore focus on sample regions containing native atomic defects, which act as effective scattering centers.
In typical topographic images of the K-cleavage plane, native defects can be categorized into four common forms: large bright protrusions (large defects), arrow-shaped defects consisting of paired protrusions, 2×1 reconstructed chain-like structures, and vacancy-like depressions (see Fig. S13). Upon closer examination, the arrow-like defects exhibit structures aligned along four orthogonal orientations, as indicated by the white-dashed arrows in Fig.~\ref{fig1}D. The QPI data analyzed in the main text were acquired from regions dominated by these arrow-like defects, where the defects in the four orientations are distributed in roughly equal numbers. Calculations reproduce the arrow-like contrast by superposing an anisotropic impurity-induced modulation of the V-$d_{xz}/d_{yz}$ states on the reconstructed K-lattice background (see Fig.~S9).
At low temperature, we measured the d$I$/d$V$ spectra using both a normal W-tip and a topological SmB$_6$-tip.
The W-tip provides access to the total local density of states (LDOS), while the SmB$_6$-tip enables an novel spin-sensitive tunneling process~\cite{aishwarya2022spin} (see Figs.  S14 and S15 for tip fabrication and calibration).
Both d$I$/d$V$ spectra exhibit a similar SDW-related suppression of spectral weight between approximately $-15$ and $+25$ meV (Fig.~\ref{fig1}E). The finite residual conductance indicates a soft, particle--hole-asymmetric reconstruction gap, allowing residual low-energy states to participate in impurity scattering.
The agreement between the two tips confirms that both measurements reflect the intrinsic electronic properties of \KV. From the STS spectra, we extracted an SDW gap size of approximately 40~meV. 
The slightly smaller gap observed with the \SM-nanowire tip,  arising from additional LDOS around 10~meV, is due to the  topological surface states of \SM.

\begin{figure}[h]
\includegraphics[width=1.0\columnwidth]{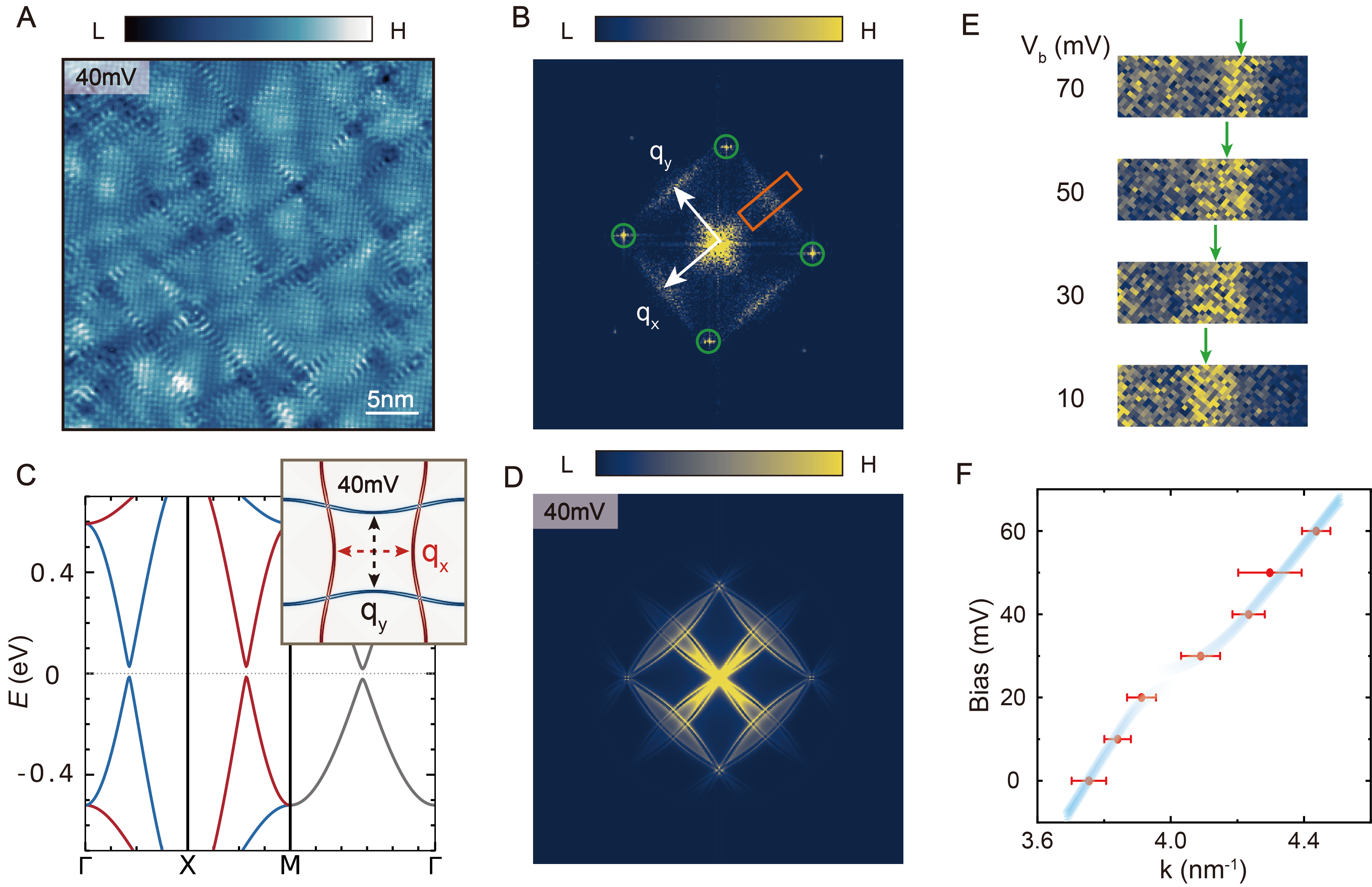}

\caption{ \textbf{Band structure of \KV.} 
\textbf{(A)} Differential conductance (d$I$/d$V$) map measuring $35 \times 35$ nm$^2$, obtained at a bias of 40~mV ($V_{\text{mod}} = 2$ mV). 
\textbf{(B)} FFT of the map in \textbf{(A)}, revealing the dominant scattering vector $q_x$ and $q_y$ (indicated by the white arrow). Green circles mark the Bragg peaks corresponding to the reconstruction lattice. The weak vertical and horizontal streaks are residual finite-field-of-view Fourier transform.
\textbf{(C)} Calculated band structure of \KV. The inset shows CEC at 40~mV. The scattering geometries for the $q_x$ and $q_y$ components of $\mathbf{q}_1$ are indicated. Note that the Fermi surface includes the first and second Brillouin zones after folding, and for clarity only $d_{xz}$ and $d_{yz}$ orbitals of  V are illustrated. 
\textbf{(D)} Theoretically calculated QPI pattern at 40~mV.  
\textbf{(E)} Evolution of the QPI signal slices within the orange rectangular region in \textbf{(B)} at various bias voltages.  
\textbf{(F)} Energy dispersion of $\mathbf{q}_x$. k is extracted from the scattering vector $q$ via the relation $q=2k$. Red dots represent peak positions extracted from Gaussian fitting, with error bars indicating the confidence intervals. The blue lines are guide lines. 
}
\label{fig2}
\end{figure}

We next measure the electronic band structure of \KV\ experimentally and compare it with the theoretical prediction of AM spin-splitting.
To achieve this, we employ the Fourier transform of quasi-particle interference (QPI) patterns acquired with a W tip, which offers high momentum-space resolution. 
Figure~\ref{fig2}A shows a representative d$I$/d$V$ map of the reconstructed $\sqrt{2}a \times \sqrt{2}a$ K surface, acquired using a W-tip at $T =$ 5~K with a bias voltage of $V_b = 40$ mV.
The defects generate pronounced Friedel oscillations with a cross-shaped pattern, which extend several nanometers along the $x$ and $y$ directions. A fast Fourier transform (FFT) of Fig.~\ref{fig2}A reveals the QPI patterns associated with multiple scattering wavevectors (see Fig.~\ref{fig2}B). By further optimizing the FFT contrast, an additional pair of peaks can also be resolved, indicative of a coexisting long-range one-dimensional electronic stripe order (Fig.  S16). Here we focus on the dominant dispersive QPI features.
The flat segments along $q_x$ originate from intra-band scatterings between the Fermi surface sections near $k_{x}\approx\pm\pi/2a$.

To understand the origin of these scattering vectors, we perform theoretical modeling and simulations of the QPI pattern (see SM for details).
Figure~\ref{fig2}C shows the band structure near the Fermi level  in which AM produces the spin splitting whereas SDW folds the bands and opens the low-energy gap; the SDW-only limit remains spin degenerate (see SM Fig.~S2).
The resulting constant energy contours (CEC) at $40$~meV (inset of Fig.~\ref{fig2}C) comprises four nearly parallel bands along the $k_x \approx \pm \pi/2a$ and $k_y \approx \pm \pi/2a$ directions. 
These orthogonal Fermi-surface segments retain opposite spin polarization and satisfy the symmetry $[U_s||M_{\bar{1}10}]$ for the spin configuration shown in Fig.~\ref{fig1}C, confirming that AM character persists even in the presence of SDW order. Because the SDW gap ($\sim 40$ meV) is substantially smaller than the AM spin-splitting ($\sim$1.8 eV), the system predominantly maintains its $d$-wave AM character. We then simulate the QPI signal by computing the Fourier-transformed LDOS modulations from elastic scattering off a single point impurity. The simulated QPI pattern at the experimental bias voltage ($V_b = 40$ mV)  reproduces the dominant scattering vectors observed near the Brillouin zone edges (see Fig.~\ref{fig2}D), with good qualitative agreement on the principal $q_x$ and $q_y$ peaks.

We also measure the dispersion of the spin-split AM bands, which is experimentally accessible through the bias voltage dependence of the scattering vectors. For example, the dispersion along the $k_x$ direction for the Fermi surface segment around $k_x = \pi/2a$ (red curves in Fig.~\ref{fig2}C) can be extracted from the energy dependence of $q_x$, which is presented in Figure~\ref{fig2}E.
Due to the SDW gap opening, the dispersion extracted from $q_x$ exhibits a  slope change between 20 and 30~meV, close to the upper edge of the SDW gap (see Fig.~\ref{fig2}F). The extended analysis from $-100$ to $+100$~mV shows no comparably sharp feature on the negative-bias side within the experimental resolution, consistent with the particle--hole asymmetry (See Fig. S17).
We next examine whether the impurity-induced standing waves carry spin-sensitive phase contrast, first in real space around individual arrow-shaped impurities and then statistically in momentum space.

\begin{figure}[h]
\includegraphics[width=1.0\columnwidth]{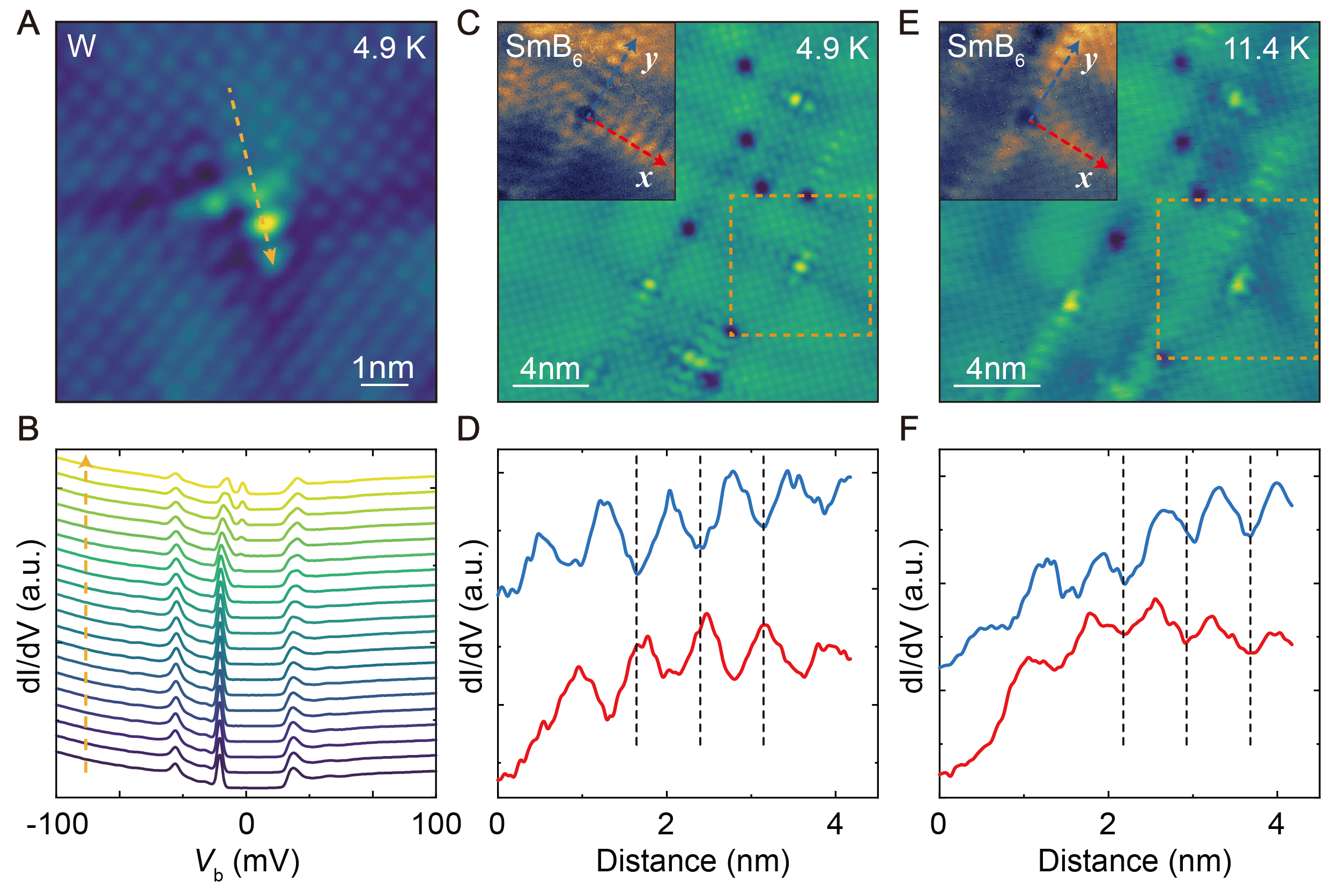}
\caption{\textbf{Spin polarized scattering in real space.} 
\textbf{(A)} Topography of K-terminated surface with a single impurity. d$I$/d$V$ line-cut along the orang dashed arrow is presented in \textbf{(B)}.
\textbf{(C, E)}  Atomical-resolved topography acquired in the same field of view using the same SmB$_6$-tip at temperatures of 4.9 K (C) and 11.4 K (E), respectively. The inset shows the corresponding $dI/dV$ map acquired from the region marked by the yellow dashed rectangle.
\textbf{(D, F)} Line cut of the d$I$/d$V$ map along  two orthogonal directions indicated by the dashed line in the inset of panel \textbf{(C)} and \textbf{(E)}.  All the curves are color coded. The line profiles are vertically offset for clarity and are intended primarily to compare the phase relation, rather than the absolute conductance amplitude. 
STM setup conditions: (A) V$_b$ = 100 mV, I$_t$ = 50 pA; (C) V$_b$ = 80 mV, I$_t$ = 40 pA, V$_{mod}$ = 2 mV; (E) V$_b$ = 80 mV, I$_t$ = 50 pA, V$_{mod}$ = 4 mV.
}
\label{fig3}
\end{figure}

Before doing so, we verify that the presence of impurities does not perturb the underlying electronic orders. To this end, we acquire d$I$/d$V$ line cuts across an impurity (see Figs.~\ref{fig3}A and B).
The results indicate that the SDW gap itself is largely unaffected by the impurity, as is the AM order. 
We therefore analyze the impurity-driven QPI patterns as probes of the host electronic structure.
However, a small peak appears around $V_b=-3$ mV, signaling the presence of in-gap bound states trapped by the lattice defect—although these states are not the focus of the present study.

Now, we examine an arrow-shaped impurity measured with the same SmB$_6$-nanowire tip in the same field of view. At 4.9~K, the standing waves extracted from the local d$I$/d$V$ map show a clear relative $\pi$ phase shift between the two orthogonal directions (Fig.~\ref{fig3}C, D). After warming to 11.4~K, the phase offset is strongly suppressed while the line-cut trajectories remain unchanged (Fig.~\ref{fig3}E, F). This same-position temperature comparison argues against a purely static impurity-shape effect or a fixed tip-frame geometrical anisotropy as the dominant origin of the low-temperature phase contrast. Importantly, 11.4~K remains far below the SDW-related transition temperature $T_N'$. The disappearance of the phase contrast while the SDW reconstruction remains present therefore disfavors a purely static SDW origin. As an independent check on the latter possibility, the same SmB$_6$-tip calibration on Fe$_{1+x}$Te resolves two orthogonal AFM domains in one field of view, showing that the spin-sensitive contrast follows the sample magnetic-domain orientation rather than a fixed tip direction (Fig. S18). The phase shift is reproducible around multiple arrow-shaped impurities measured with SmB$_6$-tip, where the oscillations along the impurity-tail direction and the perpendicular direction show a near-$\pi$ phase shift (Fig. S19). To keep consistency, the phase is limited to the directions along and perpendicular to the ``tail" of the arrow. By contrast, W-tip measurements mostly show in-phase oscillations along the two directions; when the Friedel oscillations are weak, the topmost K-lattice modulation can introduce some phase offset (see statistics in Fig. S20). 

This asymmetric response is reminiscent of spin-polarized  stripes observed in Fe(Se,Te) using SmB$_6$-nanowire tips~\cite{aishwarya2022spin}.
In that system, the measured LDOS oscillates between spin-up and spin-down atomic sites (real space). 
Moreover, these oscillations reverse phase upon reversing the bias voltage, leading to a $\pi$ phase shift in the topography along the same line cut between opposite bias voltages. We attributed this behavior to spin-selective tunneling of the SmB$_6$-tip, while the underlying tunneling mechanism is not resolved yet~\cite{banerjee_axionic_2025}. 
Furthermore, the real-space AM calculations provide a microscopic consistency check: after the K-lattice background is included, the spin-sensitive observable exhibits a near-$\pi$ phase relation between the orthogonal standing waves, whereas the spin-summed response is predominantly in phase (see Fig. S11). In addition, symmetry-preserving magnetic and non-magnetic impurity potentials produce nearly identical charge-channel $q_x/q_y$ QPI patterns, and the calculated spin-flip contribution is weak (see Fig.~S8).
Therefore, the $\pi$ phase shift  revealed by \SM-tip at low temperature  strongly indicates that $d_{xz}$ and $d_{yz}$ bands in \KV\ are oppositely spin-polarized in momentum space.

While the phase relation of the Friedel oscillations provides a local real-space signature, their amplitudes can depend on the local scattering potential and environment of individual defects. We therefore turn to the statistically averaged QPI intensity to assess the $q_x/q_y$ anisotropy more systematically.
If the  Friedel oscillations are indeed oppositely spin-polarized between $q_x$ and $q_y$ directions,  it should be readily accessible with the spin-polarized SmB$_6$-tip. Most importantly, the anisotropy between $q_x$ and $q_y$ should reverse upon reversing the bias voltage---a hallmark of spin-polarized tunneling. 
 With the W-tip, scattering from different impurities is detected without spin selectivity, so the  averaged QPI intensities should be similar between $\pm q_x$ and $\pm q_y$. Furthermore, even if the W-tip itself possesses an intrinsic anisotropy, such an anisotropy would not reverse with bias voltage, distinguishing it from the spin-driven signal we seek.

\begin{figure}[h]
\includegraphics[width=1.0\columnwidth]{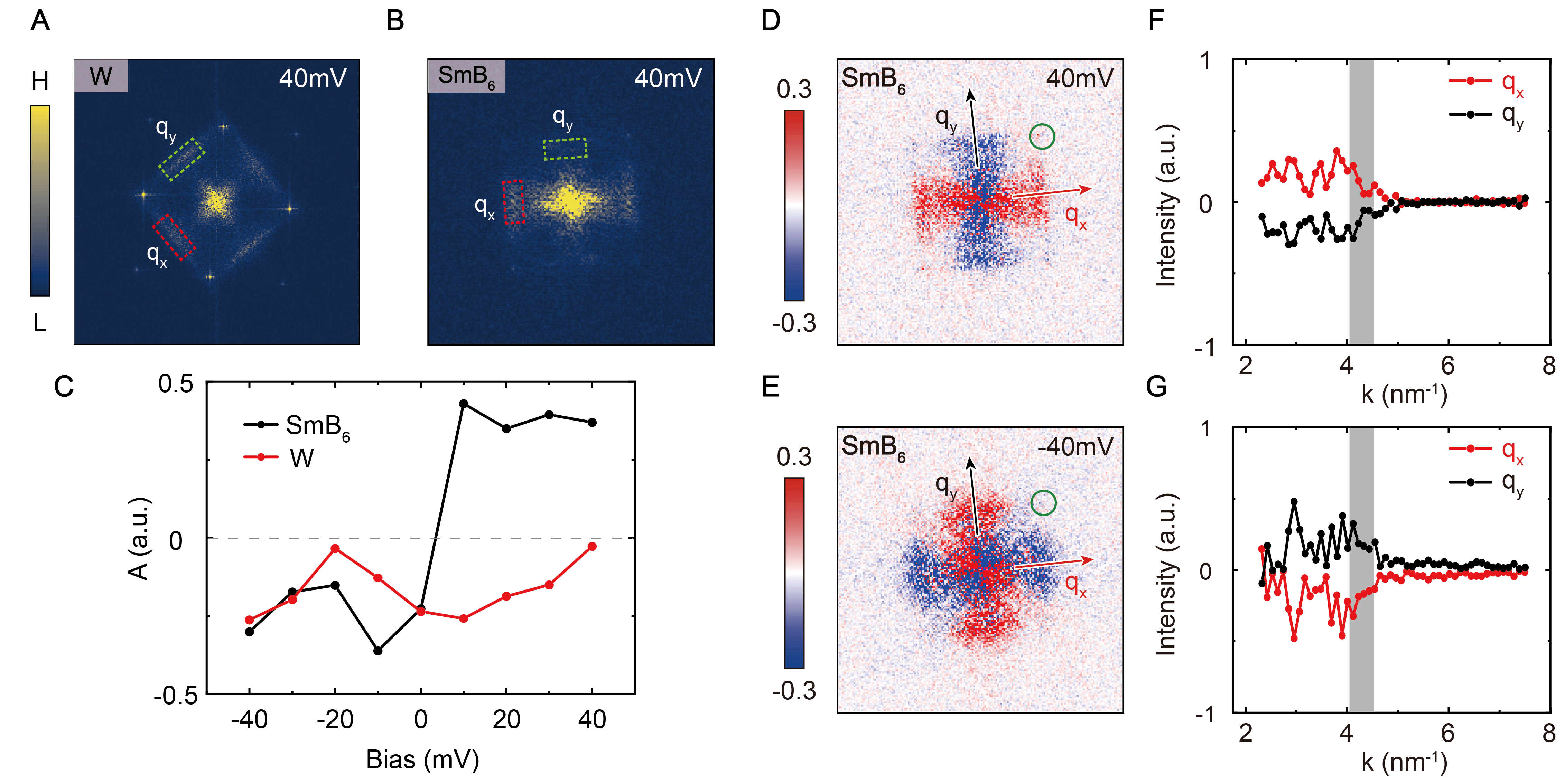}

\caption{\textbf{Visualize spin polarized band in momentum space.} \textbf{(A, B)} Fourier-transform maps of the $dI/dV$ images measured at V$_b$ = 40 mV using a conventional W-tip \textbf{(A)} and an SmB$_6$-tip \textbf{(B)}, respectively. The red and green dashed rectangles mark the regions of interest (ROI) used to integrate the raw FFT intensities around the $q_x$ and $q_y$ scattering channels. \textbf{(C)}  Energy dependence of the normalized  raw-FFT anisotropy coefficient $A(E)$. \textbf{(D, E)} Representative FFT difference maps obtained using an SmB$_6$-tip at  $\pm40$ mV. \textbf{(F, G)} Corresponding line profiles extracted along the $q_x$ and $q_y$ directions are presented in \textbf{(D)} and \textbf{(E)}. The gray shaded regions indicate the position of the scattering vector $q$ in k space at the corresponding energy.
} 
\label{fig4}
\end{figure}

To get statistical evidence for $d$-wave altermagnetism, we extended our QPI analysis to a larger field of view containing multiple impurities.
Figure~\ref{fig4} presents a comprehensive comparison between measurements acquired with W- and SmB$_6$-tips at both positive and negative bias voltages. 
We first analyze the raw FFT intensities without rotational subtraction. As shown in Figs.~\ref{fig4}A and B, we select rectangular ROIs around the $q_x$ and $q_y$ scattering channels, and then integrate over these regions to obtain $I_{q_x}(E)$ and $I_{q_y}(E)$. The normalized anisotropy coefficient is then defined as $A(E)=[I_{q_x}(E)-I_{q_y}(E)]/[I_{q_x}(E)+I_{q_y}(E)]$. This quantity quantifies the anisotropy between orthogonal directions from the raw FFT data.

For the W-tip, $A(E)$ is finite, demonstrating that spin-insensitive QPI can contain a static background anisotropy. This bias-even component may include contributions from the one-dimensional SDW reconstruction, impurity and sample-orbital matrix elements, or a residual tip-apex anisotropy. Crucially, however, it does not change sign upon reversing the bias.  
 In stark contrast, measurements with the SmB$_6$-tip yield a clear, bias-dependent anisotropy: $A(E)$ changes sign upon reversing the bias polarity, indicating that the $q_x$- and $q_y$-channel intensities are interchanged in the raw FFT data. Calculations comparing spin-sensitive tunneling and orbital-imbalanced impurity scattering show that both mechanisms can generate $C_4$-breaking QPI; however, only the spin-sensitive channel produces the observed sign reversal of $A(E)$ (see Fig.~S7). Moreover, the representative difference maps (defined in Fig. S21) at $\pm40$~mV provide a complementary visualization, and the corresponding line profiles highlight the opposite evolution of the $q_x$ and $q_y$ scattering channels as shown in Figs.~\ref{fig4}D--G. The corresponding real-space spectroscopic maps and raw FFTs for the SmB$_6$- and W-tip datasets are provided in Figs. S22 and S23, respectively. 

Taken together, our understanding of \KV does not rely on QPI anisotropy alone. The W-tip establishes a non-reversing spin-averaged background; the SmB$_6$-tip reveals an additional bias-odd raw-FFT component; the same-impurity phase contrast is suppressed at 11.4 K while the SDW state remains present; and SmB$_6$-tip calibration shows that the directional contrast follows sample magnetic domains rather than a fixed tip axis. Moreover, numerical calculations further tested SDW-only and orbital-imbalanced scenarios which cannot capture our core observations.


Because STM is predominantly surface sensitive, the present conclusion concerns the topmost V$_2$O-derived electronic states.
In the bulk, KV$_2$Se$_2$O was found to form a G-type antiferromagnetic order below $T_N$, in which the N\'eel vector alternates between adjacent layers~\cite{yang2025observation}. This magnetic structure has recently been identified as a hidden altermagnetism, characterized by an additional symmetry operation $[U_s|M_z]$, where $M_z$ is the out-of-plane mirror reflection. 
Hence, each individual layer possesses an AM character, leading to an even-odd layer dependence in few-layer samples~\cite{li2026JJevenodd}.  Nevertheless, interlayer hybridization along the $c$-axis is extremely weak, rendering the system quasi-two-dimensional. Since STM is a highly surface-sensitive probe, only the topmost layer of KV$_2$Se$_2$O contributes to our measurements, effectively realizing the monolayer AM behavior.
On the other hand, our work provides a method for investigating AM in real space. The SmB$_6$-nanowire probe not only delivers high spin resolution but also enables visualization of surface properties at the sub-nanometer scale. This capability allows us to resolve the spin structure within a single magnetic domain—a critical advantage over optical-based techniques, which average over macroscopic areas. Therefore, our SmB$_6$-tip-based STM is applicable to detect the hidden AM even–odd effect through local spin-resolved spectroscopy. More broadly, this approach could be extended to probe magnetically compensated spin textures in a wide range of correlated and topological quantum materials.

In summary, combining the normal metallic W-tip and topological \SM-nanowire tip, we provide  real-space and momentum space evidence for the AM spin splitting of \KV. 
Detailed information related to both the SDW gap and the $d$-wave type AM spin-splitting band structures are directly visualized. In addition, theoretical simulations based on a minimal $d$-wave AM model agree  well with the QPI measurements. Looking forward, the unique properties of KV$_2$Se$_2$O make it a promising candidate for spintronics applications, such as enabling efficient spin current generation without net magnetization, tunable magnetic tunnel junctions with even-odd layer-dependent responses, and ultra-low dissipation magnonic devices. Moreover, its van der Waals layered structure and robust AM order, combined with strong correlation physics, position KV$_2$Se$_2$O as a versatile building block for ultra-compact spintronic devices, magnetic tunnel junctions, and quantum sensing technologies.

\textit{Note added.--}
In the preparation of our manuscript, we become aware of two other STM work with similar results ~\cite{wang2025atomic,fu2025atomic}.

\section*{acknowledgments}

We thank Vidya Madhavan, Zhenyu Wang, Yang Liu for valuable discussions. This work was supported by the National Key R\&D Program of China (Grant No. 2022YFA1402200, 2023YFA1406100, 2024YFA1207800, and 2025YFA1411501), the National Natural Science Foundation of China (Grants Nos. 12374151, 12350710785, W2511006, 12561160109, 12574148), and the Zhejiang Provincial Natural Science Foundation of China (Grant No. LR25A040003).



\begin{thebibliography}{53}%
\makeatletter
\providecommand \@ifxundefined [1]{%
 \@ifx{#1\undefined}
}%
\providecommand \@ifnum [1]{%
 \ifnum #1\expandafter \@firstoftwo
 \else \expandafter \@secondoftwo
 \fi
}%
\providecommand \@ifx [1]{%
 \ifx #1\expandafter \@firstoftwo
 \else \expandafter \@secondoftwo
 \fi
}%
\providecommand \natexlab [1]{#1}%
\providecommand \enquote  [1]{``#1''}%
\providecommand \bibnamefont  [1]{#1}%
\providecommand \bibfnamefont [1]{#1}%
\providecommand \citenamefont [1]{#1}%
\providecommand \href@noop [0]{\@secondoftwo}%
\providecommand \href [0]{\begingroup \@sanitize@url \@href}%
\providecommand \@href[1]{\@@startlink{#1}\@@href}%
\providecommand \@@href[1]{\endgroup#1\@@endlink}%
\providecommand \@sanitize@url [0]{\catcode `\\12\catcode `\$12\catcode `\&12\catcode `\#12\catcode `\^12\catcode `\_12\catcode `\%12\relax}%
\providecommand \@@startlink[1]{}%
\providecommand \@@endlink[0]{}%
\providecommand \url  [0]{\begingroup\@sanitize@url \@url }%
\providecommand \@url [1]{\endgroup\@href {#1}{\urlprefix }}%
\providecommand \urlprefix  [0]{URL }%
\providecommand \Eprint [0]{\href }%
\providecommand \doibase [0]{https://doi.org/}%
\providecommand \selectlanguage [0]{\@gobble}%
\providecommand \bibinfo  [0]{\@secondoftwo}%
\providecommand \bibfield  [0]{\@secondoftwo}%
\providecommand \translation [1]{[#1]}%
\providecommand \BibitemOpen [0]{}%
\providecommand \bibitemStop [0]{}%
\providecommand \bibitemNoStop [0]{.\EOS\space}%
\providecommand \EOS [0]{\spacefactor3000\relax}%
\providecommand \BibitemShut  [1]{\csname bibitem#1\endcsname}%
\let\auto@bib@innerbib\@empty
\bibitem [{\citenamefont {{\v{S}}mejkal}\ \emph {et~al.}(2020)\citenamefont {{\v{S}}mejkal}, \citenamefont {Gonz{\'a}lez-Hern{\'a}ndez}, \citenamefont {Jungwirth},\ and\ \citenamefont {Sinova}}]{vsmejkal2020crystal}%
  \BibitemOpen
  \bibfield  {author} {\bibinfo {author} {\bibfnamefont {L.}~\bibnamefont {{\v{S}}mejkal}}, \bibinfo {author} {\bibfnamefont {R.}~\bibnamefont {Gonz{\'a}lez-Hern{\'a}ndez}}, \bibinfo {author} {\bibfnamefont {T.}~\bibnamefont {Jungwirth}},\ and\ \bibinfo {author} {\bibfnamefont {J.}~\bibnamefont {Sinova}},\ }\bibfield  {title} {\bibinfo {title} {{Crystal time-reversal symmetry breaking and spontaneous {Hall} effect in collinear antiferromagnets}},\ }\href {https://www.science.org/doi/full/10.1126/sciadv.aaz8809} {\bibfield  {journal} {\bibinfo  {journal} {Sci. Adv.}\ }\textbf {\bibinfo {volume} {6}},\ \bibinfo {pages} {eaaz8809} (\bibinfo {year} {2020})}\BibitemShut {NoStop}%
\bibitem [{\citenamefont {Naka}\ \emph {et~al.}(2019)\citenamefont {Naka}, \citenamefont {Hayami}, \citenamefont {Kusunose}, \citenamefont {Yanagi}, \citenamefont {Motome},\ and\ \citenamefont {Seo}}]{Naka19NC}%
  \BibitemOpen
  \bibfield  {author} {\bibinfo {author} {\bibfnamefont {M.}~\bibnamefont {Naka}}, \bibinfo {author} {\bibfnamefont {S.}~\bibnamefont {Hayami}}, \bibinfo {author} {\bibfnamefont {H.}~\bibnamefont {Kusunose}}, \bibinfo {author} {\bibfnamefont {Y.}~\bibnamefont {Yanagi}}, \bibinfo {author} {\bibfnamefont {Y.}~\bibnamefont {Motome}},\ and\ \bibinfo {author} {\bibfnamefont {H.}~\bibnamefont {Seo}},\ }\bibfield  {title} {\bibinfo {title} {Spin current generation in organic antiferromagnets},\ }\href {https://doi.org/10.1038/s41467-019-12229-y} {\bibfield  {journal} {\bibinfo  {journal} {Nat. Commun.}\ }\textbf {\bibinfo {volume} {10}},\ \bibinfo {pages} {4305} (\bibinfo {year} {2019})}\BibitemShut {NoStop}%
\bibitem [{\citenamefont {Hayami}\ \emph {et~al.}(2019)\citenamefont {Hayami}, \citenamefont {Yanagi},\ and\ \citenamefont {Kusunose}}]{hayami2019momentum}%
  \BibitemOpen
  \bibfield  {author} {\bibinfo {author} {\bibfnamefont {S.}~\bibnamefont {Hayami}}, \bibinfo {author} {\bibfnamefont {Y.}~\bibnamefont {Yanagi}},\ and\ \bibinfo {author} {\bibfnamefont {H.}~\bibnamefont {Kusunose}},\ }\bibfield  {title} {\bibinfo {title} {Momentum-dependent spin splitting by collinear antiferromagnetic ordering},\ }\href {https://doi.org/10.7566/JPSJ.88.123702} {\bibfield  {journal} {\bibinfo  {journal} {J. Phys. Soc. Jpn.}\ }\textbf {\bibinfo {volume} {88}},\ \bibinfo {pages} {123702} (\bibinfo {year} {2019})}\BibitemShut {NoStop}%
\bibitem [{\citenamefont {\ifmmode~\check{S}\else \v{S}\fi{}mejkal}\ \emph {et~al.}(2022{\natexlab{a}})\citenamefont {\ifmmode~\check{S}\else \v{S}\fi{}mejkal}, \citenamefont {Sinova},\ and\ \citenamefont {Jungwirth}}]{ifmmode2022Beyond}%
  \BibitemOpen
  \bibfield  {author} {\bibinfo {author} {\bibfnamefont {L.}~\bibnamefont {\ifmmode~\check{S}\else \v{S}\fi{}mejkal}}, \bibinfo {author} {\bibfnamefont {J.}~\bibnamefont {Sinova}},\ and\ \bibinfo {author} {\bibfnamefont {T.}~\bibnamefont {Jungwirth}},\ }\bibfield  {title} {\bibinfo {title} {Beyond conventional ferromagnetism and antiferromagnetism: A phase with nonrelativistic spin and crystal rotation symmetry},\ }\href {https://doi.org/10.1103/PhysRevX.12.031042} {\bibfield  {journal} {\bibinfo  {journal} {Phys. Rev. X}\ }\textbf {\bibinfo {volume} {12}},\ \bibinfo {pages} {031042} (\bibinfo {year} {2022}{\natexlab{a}})}\BibitemShut {NoStop}%
\bibitem [{\citenamefont {Jungwirth}\ \emph {et~al.}(2025)\citenamefont {Jungwirth}, \citenamefont {Fernandes}, \citenamefont {Fradkin}, \citenamefont {MacDonald}, \citenamefont {Sinova},\ and\ \citenamefont {{\v{S}}mejkal}}]{jungwirth2025altermagnetism}%
  \BibitemOpen
  \bibfield  {author} {\bibinfo {author} {\bibfnamefont {T.}~\bibnamefont {Jungwirth}}, \bibinfo {author} {\bibfnamefont {R.~M.}\ \bibnamefont {Fernandes}}, \bibinfo {author} {\bibfnamefont {E.}~\bibnamefont {Fradkin}}, \bibinfo {author} {\bibfnamefont {A.~H.}\ \bibnamefont {MacDonald}}, \bibinfo {author} {\bibfnamefont {J.}~\bibnamefont {Sinova}},\ and\ \bibinfo {author} {\bibfnamefont {L.}~\bibnamefont {{\v{S}}mejkal}},\ }\bibfield  {title} {\bibinfo {title} {Altermagnetism: an unconventional spin-ordered phase of matter},\ }\href {http://dx.doi.org/10.1016/j.newton.2025.100162} {\bibfield  {journal} {\bibinfo  {journal} {Newton}\ }\textbf {\bibinfo {volume} {1}},\ \bibinfo {pages} {100162} (\bibinfo {year} {2025})}\BibitemShut {NoStop}%
\bibitem [{\citenamefont {Song}\ \emph {et~al.}(2025)\citenamefont {Song}, \citenamefont {Bai}, \citenamefont {Zhou}, \citenamefont {Han}, \citenamefont {Reichlova}, \citenamefont {Dil}, \citenamefont {Liu}, \citenamefont {Chen},\ and\ \citenamefont {Pan}}]{song2025altermagnets}%
  \BibitemOpen
  \bibfield  {author} {\bibinfo {author} {\bibfnamefont {C.}~\bibnamefont {Song}}, \bibinfo {author} {\bibfnamefont {H.}~\bibnamefont {Bai}}, \bibinfo {author} {\bibfnamefont {Z.}~\bibnamefont {Zhou}}, \bibinfo {author} {\bibfnamefont {L.}~\bibnamefont {Han}}, \bibinfo {author} {\bibfnamefont {H.}~\bibnamefont {Reichlova}}, \bibinfo {author} {\bibfnamefont {J.~H.}\ \bibnamefont {Dil}}, \bibinfo {author} {\bibfnamefont {J.}~\bibnamefont {Liu}}, \bibinfo {author} {\bibfnamefont {X.}~\bibnamefont {Chen}},\ and\ \bibinfo {author} {\bibfnamefont {F.}~\bibnamefont {Pan}},\ }\bibfield  {title} {\bibinfo {title} {Altermagnets as a new class of functional materials},\ }\href {https://doi.org/10.1038/s41578-025-00779-1} {\bibfield  {journal} {\bibinfo  {journal} {Nat. Rev. Mater.}\ }\textbf {\bibinfo {volume} {10}},\ \bibinfo {pages} {473} (\bibinfo {year} {2025})}\BibitemShut {NoStop}%
\bibitem [{\citenamefont {Jungwirth}\ \emph {et~al.}(2026)\citenamefont {Jungwirth}, \citenamefont {Sinova}, \citenamefont {Fernandes}, \citenamefont {Liu}, \citenamefont {Watanabe}, \citenamefont {Murakami}, \citenamefont {Nakatsuji},\ and\ \citenamefont {{\v{S}}mejkal}}]{jungwirth2026symmetry}%
  \BibitemOpen
  \bibfield  {author} {\bibinfo {author} {\bibfnamefont {T.}~\bibnamefont {Jungwirth}}, \bibinfo {author} {\bibfnamefont {J.}~\bibnamefont {Sinova}}, \bibinfo {author} {\bibfnamefont {R.~M.}\ \bibnamefont {Fernandes}}, \bibinfo {author} {\bibfnamefont {Q.}~\bibnamefont {Liu}}, \bibinfo {author} {\bibfnamefont {H.}~\bibnamefont {Watanabe}}, \bibinfo {author} {\bibfnamefont {S.}~\bibnamefont {Murakami}}, \bibinfo {author} {\bibfnamefont {S.}~\bibnamefont {Nakatsuji}},\ and\ \bibinfo {author} {\bibfnamefont {L.}~\bibnamefont {{\v{S}}mejkal}},\ }\bibfield  {title} {\bibinfo {title} {Symmetry, microscopy and spectroscopy signatures of altermagnetism},\ }\href {https://doi.org/10.1038/s41586-025-09883-2} {\bibfield  {journal} {\bibinfo  {journal} {Nature}\ }\textbf {\bibinfo {volume} {649}},\ \bibinfo {pages} {837} (\bibinfo {year} {2026})}\BibitemShut {NoStop}%
\bibitem [{\citenamefont {Hirsch}(1990)}]{Hirsch1990prb}%
  \BibitemOpen
  \bibfield  {author} {\bibinfo {author} {\bibfnamefont {J.~E.}\ \bibnamefont {Hirsch}},\ }\bibfield  {title} {\bibinfo {title} {Spin-split states in metals},\ }\href {https://doi.org/10.1103/PhysRevB.41.6820} {\bibfield  {journal} {\bibinfo  {journal} {Phys. Rev. B}\ }\textbf {\bibinfo {volume} {41}},\ \bibinfo {pages} {6820} (\bibinfo {year} {1990})}\BibitemShut {NoStop}%
\bibitem [{\citenamefont {Ikeda}\ and\ \citenamefont {Ohashi}(1998)}]{Ikeda1998prl}%
  \BibitemOpen
  \bibfield  {author} {\bibinfo {author} {\bibfnamefont {H.}~\bibnamefont {Ikeda}}\ and\ \bibinfo {author} {\bibfnamefont {Y.}~\bibnamefont {Ohashi}},\ }\bibfield  {title} {\bibinfo {title} {Theory of unconventional spin density wave: A possible mechanism of the micromagnetism in {U-based} heavy fermion compounds},\ }\href {https://doi.org/10.1103/PhysRevLett.81.3723} {\bibfield  {journal} {\bibinfo  {journal} {Phys. Rev. Lett.}\ }\textbf {\bibinfo {volume} {81}},\ \bibinfo {pages} {3723} (\bibinfo {year} {1998})}\BibitemShut {NoStop}%
\bibitem [{\citenamefont {Wu}\ and\ \citenamefont {Zhang}(2004)}]{wu2004prl}%
  \BibitemOpen
  \bibfield  {author} {\bibinfo {author} {\bibfnamefont {C.}~\bibnamefont {Wu}}\ and\ \bibinfo {author} {\bibfnamefont {S.-C.}\ \bibnamefont {Zhang}},\ }\bibfield  {title} {\bibinfo {title} {Dynamic generation of spin-orbit coupling},\ }\href {https://doi.org/10.1103/PhysRevLett.93.036403} {\bibfield  {journal} {\bibinfo  {journal} {Phys. Rev. Lett.}\ }\textbf {\bibinfo {volume} {93}},\ \bibinfo {pages} {036403} (\bibinfo {year} {2004})}\BibitemShut {NoStop}%
\bibitem [{\citenamefont {Wu}\ \emph {et~al.}(2007)\citenamefont {Wu}, \citenamefont {Sun}, \citenamefont {Fradkin},\ and\ \citenamefont {Zhang}}]{CJWu07PRB}%
  \BibitemOpen
  \bibfield  {author} {\bibinfo {author} {\bibfnamefont {C.}~\bibnamefont {Wu}}, \bibinfo {author} {\bibfnamefont {K.}~\bibnamefont {Sun}}, \bibinfo {author} {\bibfnamefont {E.}~\bibnamefont {Fradkin}},\ and\ \bibinfo {author} {\bibfnamefont {S.-C.}\ \bibnamefont {Zhang}},\ }\bibfield  {title} {\bibinfo {title} {Fermi liquid instabilities in the spin channel},\ }\href {https://doi.org/10.1103/PhysRevB.75.115103} {\bibfield  {journal} {\bibinfo  {journal} {Phys. Rev. B}\ }\textbf {\bibinfo {volume} {75}},\ \bibinfo {pages} {115103} (\bibinfo {year} {2007})}\BibitemShut {NoStop}%
\bibitem [{\citenamefont {Chen}\ \emph {et~al.}(2014)\citenamefont {Chen}, \citenamefont {Niu},\ and\ \citenamefont {MacDonald}}]{Chen2014Anomalous}%
  \BibitemOpen
  \bibfield  {author} {\bibinfo {author} {\bibfnamefont {H.}~\bibnamefont {Chen}}, \bibinfo {author} {\bibfnamefont {Q.}~\bibnamefont {Niu}},\ and\ \bibinfo {author} {\bibfnamefont {A.~H.}\ \bibnamefont {MacDonald}},\ }\bibfield  {title} {\bibinfo {title} {Anomalous {Hall} effect arising from noncollinear antiferromagnetism},\ }\href {https://doi.org/10.1103/PhysRevLett.112.017205} {\bibfield  {journal} {\bibinfo  {journal} {Phys. Rev. Lett.}\ }\textbf {\bibinfo {volume} {112}},\ \bibinfo {pages} {017205} (\bibinfo {year} {2014})}\BibitemShut {NoStop}%
\bibitem [{\citenamefont {Liu}\ \emph {et~al.}(2025{\natexlab{a}})\citenamefont {Liu}, \citenamefont {Dai},\ and\ \citenamefont {Bl{\"u}gel}}]{liu2025different}%
  \BibitemOpen
  \bibfield  {author} {\bibinfo {author} {\bibfnamefont {Q.}~\bibnamefont {Liu}}, \bibinfo {author} {\bibfnamefont {X.}~\bibnamefont {Dai}},\ and\ \bibinfo {author} {\bibfnamefont {S.}~\bibnamefont {Bl{\"u}gel}},\ }\bibfield  {title} {\bibinfo {title} {Different facets of unconventional magnetism},\ }\href {https://doi.org/10.1038/s41567-024-02750-3} {\bibfield  {journal} {\bibinfo  {journal} {Nat. Phys.}\ }\textbf {\bibinfo {volume} {21}},\ \bibinfo {pages} {329} (\bibinfo {year} {2025}{\natexlab{a}})}\BibitemShut {NoStop}%
\bibitem [{\citenamefont {Liu}\ \emph {et~al.}(2025{\natexlab{b}})\citenamefont {Liu}, \citenamefont {Wei}, \citenamefont {Peng}, \citenamefont {Hou}, \citenamefont {Gao},\ and\ \citenamefont {Niu}}]{liu2025prx}%
  \BibitemOpen
  \bibfield  {author} {\bibinfo {author} {\bibfnamefont {Z.}~\bibnamefont {Liu}}, \bibinfo {author} {\bibfnamefont {M.}~\bibnamefont {Wei}}, \bibinfo {author} {\bibfnamefont {W.}~\bibnamefont {Peng}}, \bibinfo {author} {\bibfnamefont {D.}~\bibnamefont {Hou}}, \bibinfo {author} {\bibfnamefont {Y.}~\bibnamefont {Gao}},\ and\ \bibinfo {author} {\bibfnamefont {Q.}~\bibnamefont {Niu}},\ }\bibfield  {title} {\bibinfo {title} {Multipolar anisotropy in anomalous {Hall} effect from spin-group symmetry breaking},\ }\href {https://doi.org/10.1103/PhysRevX.15.031006} {\bibfield  {journal} {\bibinfo  {journal} {Phys. Rev. X}\ }\textbf {\bibinfo {volume} {15}},\ \bibinfo {pages} {031006} (\bibinfo {year} {2025}{\natexlab{b}})}\BibitemShut {NoStop}%
\bibitem [{\citenamefont {Liu}\ \emph {et~al.}(2025{\natexlab{c}})\citenamefont {Liu}, \citenamefont {Chen}, \citenamefont {Yu},\ and\ \citenamefont {Liu}}]{liu2025symmetry}%
  \BibitemOpen
  \bibfield  {author} {\bibinfo {author} {\bibfnamefont {Y.}~\bibnamefont {Liu}}, \bibinfo {author} {\bibfnamefont {X.}~\bibnamefont {Chen}}, \bibinfo {author} {\bibfnamefont {Y.}~\bibnamefont {Yu}},\ and\ \bibinfo {author} {\bibfnamefont {Q.}~\bibnamefont {Liu}},\ }\bibfield  {title} {\bibinfo {title} {Symmetry classification of magnetic orders and emergence of spin-orbit magnetism},\ }\href {https://doi.org/10.48550/arXiv.2506.20739} {\bibfield  {journal} {\bibinfo  {journal} {arXiv preprint arXiv:2506.20739}\ } (\bibinfo {year} {2025}{\natexlab{c}})}\BibitemShut {NoStop}%
\bibitem [{\citenamefont {\ifmmode~\check{S}\else \v{S}\fi{}mejkal}\ \emph {et~al.}(2022{\natexlab{b}})\citenamefont {\ifmmode~\check{S}\else \v{S}\fi{}mejkal}, \citenamefont {Sinova},\ and\ \citenamefont {Jungwirth}}]{ifmmode2022Emerging}%
  \BibitemOpen
  \bibfield  {author} {\bibinfo {author} {\bibfnamefont {L.}~\bibnamefont {\ifmmode~\check{S}\else \v{S}\fi{}mejkal}}, \bibinfo {author} {\bibfnamefont {J.}~\bibnamefont {Sinova}},\ and\ \bibinfo {author} {\bibfnamefont {T.}~\bibnamefont {Jungwirth}},\ }\bibfield  {title} {\bibinfo {title} {Emerging research landscape of altermagnetism},\ }\href {https://doi.org/10.1103/PhysRevX.12.040501} {\bibfield  {journal} {\bibinfo  {journal} {Phys. Rev. X}\ }\textbf {\bibinfo {volume} {12}},\ \bibinfo {pages} {040501} (\bibinfo {year} {2022}{\natexlab{b}})}\BibitemShut {NoStop}%
\bibitem [{\citenamefont {Bai}\ \emph {et~al.}(2024{\natexlab{a}})\citenamefont {Bai}, \citenamefont {Feng}, \citenamefont {Liu}, \citenamefont {{\v{S}}mejkal}, \citenamefont {Mokrousov},\ and\ \citenamefont {Yao}}]{bai2024altermagnetism}%
  \BibitemOpen
  \bibfield  {author} {\bibinfo {author} {\bibfnamefont {L.}~\bibnamefont {Bai}}, \bibinfo {author} {\bibfnamefont {W.}~\bibnamefont {Feng}}, \bibinfo {author} {\bibfnamefont {S.}~\bibnamefont {Liu}}, \bibinfo {author} {\bibfnamefont {L.}~\bibnamefont {{\v{S}}mejkal}}, \bibinfo {author} {\bibfnamefont {Y.}~\bibnamefont {Mokrousov}},\ and\ \bibinfo {author} {\bibfnamefont {Y.}~\bibnamefont {Yao}},\ }\bibfield  {title} {\bibinfo {title} {Altermagnetism: Exploring new frontiers in magnetism and spintronics},\ }\href {https://onlinelibrary.wiley.com/doi/abs/10.1002/adfm.202409327} {\bibfield  {journal} {\bibinfo  {journal} {Adv. Funct. Mater.}\ }\textbf {\bibinfo {volume} {34}},\ \bibinfo {pages} {2409327} (\bibinfo {year} {2024}{\natexlab{a}})}\BibitemShut {NoStop}%
\bibitem [{\citenamefont {Fender}\ \emph {et~al.}(2025)\citenamefont {Fender}, \citenamefont {Gonzalez},\ and\ \citenamefont {Bediako}}]{fender2025altermagnetism}%
  \BibitemOpen
  \bibfield  {author} {\bibinfo {author} {\bibfnamefont {S.~S.}\ \bibnamefont {Fender}}, \bibinfo {author} {\bibfnamefont {O.}~\bibnamefont {Gonzalez}},\ and\ \bibinfo {author} {\bibfnamefont {D.~K.}\ \bibnamefont {Bediako}},\ }\bibfield  {title} {\bibinfo {title} {Altermagnetism: A chemical perspective},\ }\href {https://pubs.acs.org/doi/abs/10.1021/jacs.4c14503} {\bibfield  {journal} {\bibinfo  {journal} {J. Am. Chem. Soc.}\ }\textbf {\bibinfo {volume} {147}},\ \bibinfo {pages} {2257} (\bibinfo {year} {2025})}\BibitemShut {NoStop}%
\bibitem [{\citenamefont {Gonz\'alez-Hern\'andez}\ \emph {et~al.}(2021)\citenamefont {Gonz\'alez-Hern\'andez}, \citenamefont {\ifmmode~\check{S}\else \v{S}\fi{}mejkal}, \citenamefont {V\'yborn\'y}, \citenamefont {Yahagi}, \citenamefont {Sinova}, \citenamefont {Jungwirth},\ and\ \citenamefont {\ifmmode~\check{Z}\else \v{Z}\fi{}elezn\'y}}]{gonzal2021ruo2prl}%
  \BibitemOpen
  \bibfield  {author} {\bibinfo {author} {\bibfnamefont {R.}~\bibnamefont {Gonz\'alez-Hern\'andez}}, \bibinfo {author} {\bibfnamefont {L.}~\bibnamefont {\ifmmode~\check{S}\else \v{S}\fi{}mejkal}}, \bibinfo {author} {\bibfnamefont {K.}~\bibnamefont {V\'yborn\'y}}, \bibinfo {author} {\bibfnamefont {Y.}~\bibnamefont {Yahagi}}, \bibinfo {author} {\bibfnamefont {J.}~\bibnamefont {Sinova}}, \bibinfo {author} {\bibfnamefont {T.~c.~v.}\ \bibnamefont {Jungwirth}},\ and\ \bibinfo {author} {\bibfnamefont {J.}~\bibnamefont {\ifmmode~\check{Z}\else \v{Z}\fi{}elezn\'y}},\ }\bibfield  {title} {\bibinfo {title} {Efficient electrical spin splitter based on nonrelativistic collinear antiferromagnetism},\ }\href {https://doi.org/10.1103/PhysRevLett.126.127701} {\bibfield  {journal} {\bibinfo  {journal} {Phys. Rev. Lett.}\ }\textbf {\bibinfo {volume} {126}},\ \bibinfo {pages} {127701} (\bibinfo {year} {2021})}\BibitemShut {NoStop}%
\bibitem [{\citenamefont {Bai}\ \emph {et~al.}(2022)\citenamefont {Bai}, \citenamefont {Han}, \citenamefont {Feng}, \citenamefont {Zhou}, \citenamefont {Su}, \citenamefont {Wang}, \citenamefont {Liao}, \citenamefont {Zhu}, \citenamefont {Chen}, \citenamefont {Pan}, \citenamefont {Fan},\ and\ \citenamefont {Song}}]{bai_observation_2022}%
  \BibitemOpen
  \bibfield  {author} {\bibinfo {author} {\bibfnamefont {H.}~\bibnamefont {Bai}}, \bibinfo {author} {\bibfnamefont {L.}~\bibnamefont {Han}}, \bibinfo {author} {\bibfnamefont {X.}~\bibnamefont {Feng}}, \bibinfo {author} {\bibfnamefont {Y.}~\bibnamefont {Zhou}}, \bibinfo {author} {\bibfnamefont {R.}~\bibnamefont {Su}}, \bibinfo {author} {\bibfnamefont {Q.}~\bibnamefont {Wang}}, \bibinfo {author} {\bibfnamefont {L.}~\bibnamefont {Liao}}, \bibinfo {author} {\bibfnamefont {W.}~\bibnamefont {Zhu}}, \bibinfo {author} {\bibfnamefont {X.}~\bibnamefont {Chen}}, \bibinfo {author} {\bibfnamefont {F.}~\bibnamefont {Pan}}, \bibinfo {author} {\bibfnamefont {X.}~\bibnamefont {Fan}},\ and\ \bibinfo {author} {\bibfnamefont {C.}~\bibnamefont {Song}},\ }\bibfield  {title} {\bibinfo {title} {Observation of spin splitting torque in a collinear antiferromagnet {RuO$_{2}$}},\ }\href {https://doi.org/10.1103/PhysRevLett.128.197202} {\bibfield  {journal} {\bibinfo  {journal} {Phys. Rev. Lett.}\ }\textbf {\bibinfo {volume} {128}},\
  \bibinfo {pages} {197202} (\bibinfo {year} {2022})}\BibitemShut {NoStop}%
\bibitem [{\citenamefont {Bose}\ \emph {et~al.}(2022)\citenamefont {Bose}, \citenamefont {Schreiber}, \citenamefont {Jain}, \citenamefont {Shao}, \citenamefont {Nair}, \citenamefont {Sun}, \citenamefont {Zhang}, \citenamefont {Muller}, \citenamefont {Tsymbal}, \citenamefont {Schlom},\ and\ \citenamefont {Ralph}}]{bose_tilted_2022}%
  \BibitemOpen
  \bibfield  {author} {\bibinfo {author} {\bibfnamefont {A.}~\bibnamefont {Bose}}, \bibinfo {author} {\bibfnamefont {N.~J.}\ \bibnamefont {Schreiber}}, \bibinfo {author} {\bibfnamefont {R.}~\bibnamefont {Jain}}, \bibinfo {author} {\bibfnamefont {D.-F.}\ \bibnamefont {Shao}}, \bibinfo {author} {\bibfnamefont {H.~P.}\ \bibnamefont {Nair}}, \bibinfo {author} {\bibfnamefont {J.}~\bibnamefont {Sun}}, \bibinfo {author} {\bibfnamefont {X.~S.}\ \bibnamefont {Zhang}}, \bibinfo {author} {\bibfnamefont {D.~A.}\ \bibnamefont {Muller}}, \bibinfo {author} {\bibfnamefont {E.~Y.}\ \bibnamefont {Tsymbal}}, \bibinfo {author} {\bibfnamefont {D.~G.}\ \bibnamefont {Schlom}},\ and\ \bibinfo {author} {\bibfnamefont {D.~C.}\ \bibnamefont {Ralph}},\ }\bibfield  {title} {\bibinfo {title} {Tilted spin current generated by the collinear antiferromagnet ruthenium dioxide},\ }\href {https://doi.org/10.1038/s41928-022-00744-8} {\bibfield  {journal} {\bibinfo  {journal} {Nat. Electron.}\ }\textbf {\bibinfo {volume} {5}},\ \bibinfo {pages}
  {267} (\bibinfo {year} {2022})}\BibitemShut {NoStop}%
\bibitem [{\citenamefont {Karube}\ \emph {et~al.}(2022)\citenamefont {Karube}, \citenamefont {Tanaka}, \citenamefont {Sugawara}, \citenamefont {Kadoguchi}, \citenamefont {Kohda},\ and\ \citenamefont {Nitta}}]{karube_observation_2022}%
  \BibitemOpen
  \bibfield  {author} {\bibinfo {author} {\bibfnamefont {S.}~\bibnamefont {Karube}}, \bibinfo {author} {\bibfnamefont {T.}~\bibnamefont {Tanaka}}, \bibinfo {author} {\bibfnamefont {D.}~\bibnamefont {Sugawara}}, \bibinfo {author} {\bibfnamefont {N.}~\bibnamefont {Kadoguchi}}, \bibinfo {author} {\bibfnamefont {M.}~\bibnamefont {Kohda}},\ and\ \bibinfo {author} {\bibfnamefont {J.}~\bibnamefont {Nitta}},\ }\bibfield  {title} {\bibinfo {title} {Observation of spin-splitter torque in collinear antiferromagnetic {RuO$_{2}$}},\ }\href {https://doi.org/10.1103/PhysRevLett.129.137201} {\bibfield  {journal} {\bibinfo  {journal} {Phys. Rev. Lett.}\ }\textbf {\bibinfo {volume} {129}},\ \bibinfo {pages} {137201} (\bibinfo {year} {2022})}\BibitemShut {NoStop}%
\bibitem [{\citenamefont {Fedchenko}\ \emph {et~al.}(2024)\citenamefont {Fedchenko}, \citenamefont {Min{\'a}r}, \citenamefont {Akashdeep}, \citenamefont {D’Souza}, \citenamefont {Vasilyev}, \citenamefont {Tkach}, \citenamefont {Odenbreit}, \citenamefont {Nguyen}, \citenamefont {Kutnyakhov}, \citenamefont {Wind} \emph {et~al.}}]{fedchenko2024ruo2}%
  \BibitemOpen
  \bibfield  {author} {\bibinfo {author} {\bibfnamefont {O.}~\bibnamefont {Fedchenko}}, \bibinfo {author} {\bibfnamefont {J.}~\bibnamefont {Min{\'a}r}}, \bibinfo {author} {\bibfnamefont {A.}~\bibnamefont {Akashdeep}}, \bibinfo {author} {\bibfnamefont {S.~W.}\ \bibnamefont {D’Souza}}, \bibinfo {author} {\bibfnamefont {D.}~\bibnamefont {Vasilyev}}, \bibinfo {author} {\bibfnamefont {O.}~\bibnamefont {Tkach}}, \bibinfo {author} {\bibfnamefont {L.}~\bibnamefont {Odenbreit}}, \bibinfo {author} {\bibfnamefont {Q.}~\bibnamefont {Nguyen}}, \bibinfo {author} {\bibfnamefont {D.}~\bibnamefont {Kutnyakhov}}, \bibinfo {author} {\bibfnamefont {N.}~\bibnamefont {Wind}}, \emph {et~al.},\ }\bibfield  {title} {\bibinfo {title} {Observation of time-reversal symmetry breaking in the band structure of altermagnetic {RuO$_{2}$}},\ }\href {https://www.science.org/doi/abs/10.1126/sciadv.adj4883} {\bibfield  {journal} {\bibinfo  {journal} {Sci. Adv.}\ }\textbf {\bibinfo {volume} {10}},\ \bibinfo {pages} {eadj4883} (\bibinfo {year}
  {2024})}\BibitemShut {NoStop}%
\bibitem [{\citenamefont {Amin}\ \emph {et~al.}(2024)\citenamefont {Amin}, \citenamefont {Dal~Din}, \citenamefont {Golias}, \citenamefont {Niu}, \citenamefont {Zakharov}, \citenamefont {Fromage}, \citenamefont {Fields}, \citenamefont {Heywood}, \citenamefont {Cousins}, \citenamefont {Maccherozzi}, \citenamefont {Krempaský}, \citenamefont {Dil}, \citenamefont {Kriegner}, \citenamefont {Kiraly}, \citenamefont {Campion}, \citenamefont {Rushforth}, \citenamefont {Edmonds}, \citenamefont {Dhesi}, \citenamefont {Šmejkal}, \citenamefont {Jungwirth},\ and\ \citenamefont {Wadley}}]{amin_nanoscale_2024}%
  \BibitemOpen
  \bibfield  {author} {\bibinfo {author} {\bibfnamefont {O.~J.}\ \bibnamefont {Amin}}, \bibinfo {author} {\bibfnamefont {A.}~\bibnamefont {Dal~Din}}, \bibinfo {author} {\bibfnamefont {E.}~\bibnamefont {Golias}}, \bibinfo {author} {\bibfnamefont {Y.}~\bibnamefont {Niu}}, \bibinfo {author} {\bibfnamefont {A.}~\bibnamefont {Zakharov}}, \bibinfo {author} {\bibfnamefont {S.~C.}\ \bibnamefont {Fromage}}, \bibinfo {author} {\bibfnamefont {C.~J.~B.}\ \bibnamefont {Fields}}, \bibinfo {author} {\bibfnamefont {S.~L.}\ \bibnamefont {Heywood}}, \bibinfo {author} {\bibfnamefont {R.~B.}\ \bibnamefont {Cousins}}, \bibinfo {author} {\bibfnamefont {F.}~\bibnamefont {Maccherozzi}}, \bibinfo {author} {\bibfnamefont {J.}~\bibnamefont {Krempaský}}, \bibinfo {author} {\bibfnamefont {J.~H.}\ \bibnamefont {Dil}}, \bibinfo {author} {\bibfnamefont {D.}~\bibnamefont {Kriegner}}, \bibinfo {author} {\bibfnamefont {B.}~\bibnamefont {Kiraly}}, \bibinfo {author} {\bibfnamefont {R.~P.}\ \bibnamefont {Campion}}, \bibinfo {author}
  {\bibfnamefont {A.~W.}\ \bibnamefont {Rushforth}}, \bibinfo {author} {\bibfnamefont {K.~W.}\ \bibnamefont {Edmonds}}, \bibinfo {author} {\bibfnamefont {S.~S.}\ \bibnamefont {Dhesi}}, \bibinfo {author} {\bibfnamefont {L.}~\bibnamefont {Šmejkal}}, \bibinfo {author} {\bibfnamefont {T.}~\bibnamefont {Jungwirth}},\ and\ \bibinfo {author} {\bibfnamefont {P.}~\bibnamefont {Wadley}},\ }\bibfield  {title} {\bibinfo {title} {Nanoscale imaging and control of altermagnetism in {MnTe}},\ }\href {https://doi.org/10.1038/s41586-024-08234-x} {\bibfield  {journal} {\bibinfo  {journal} {Nature}\ }\textbf {\bibinfo {volume} {636}},\ \bibinfo {pages} {348} (\bibinfo {year} {2024})}\BibitemShut {NoStop}%
\bibitem [{\citenamefont {Krempask{\`y}}\ \emph {et~al.}(2024)\citenamefont {Krempask{\`y}}, \citenamefont {{\v{S}}mejkal}, \citenamefont {D’souza}, \citenamefont {Hajlaoui}, \citenamefont {Springholz}, \citenamefont {Uhl{\'\i}{\v{r}}ov{\'a}}, \citenamefont {Alarab}, \citenamefont {Constantinou}, \citenamefont {Strocov}, \citenamefont {Usanov} \emph {et~al.}}]{krempasky2024altermagnetic}%
  \BibitemOpen
  \bibfield  {author} {\bibinfo {author} {\bibfnamefont {J.}~\bibnamefont {Krempask{\`y}}}, \bibinfo {author} {\bibfnamefont {L.}~\bibnamefont {{\v{S}}mejkal}}, \bibinfo {author} {\bibfnamefont {S.}~\bibnamefont {D’souza}}, \bibinfo {author} {\bibfnamefont {M.}~\bibnamefont {Hajlaoui}}, \bibinfo {author} {\bibfnamefont {G.}~\bibnamefont {Springholz}}, \bibinfo {author} {\bibfnamefont {K.}~\bibnamefont {Uhl{\'\i}{\v{r}}ov{\'a}}}, \bibinfo {author} {\bibfnamefont {F.}~\bibnamefont {Alarab}}, \bibinfo {author} {\bibfnamefont {P.}~\bibnamefont {Constantinou}}, \bibinfo {author} {\bibfnamefont {V.}~\bibnamefont {Strocov}}, \bibinfo {author} {\bibfnamefont {D.}~\bibnamefont {Usanov}}, \emph {et~al.},\ }\bibfield  {title} {\bibinfo {title} {Altermagnetic lifting of kramers spin degeneracy},\ }\href {https://doi.org/10.1038/s41586-023-06907-7} {\bibfield  {journal} {\bibinfo  {journal} {Nature}\ }\textbf {\bibinfo {volume} {626}},\ \bibinfo {pages} {517} (\bibinfo {year} {2024})}\BibitemShut {NoStop}%
\bibitem [{\citenamefont {Lee}\ \emph {et~al.}(2024)\citenamefont {Lee}, \citenamefont {Lee}, \citenamefont {Jung}, \citenamefont {Jung}, \citenamefont {Kim}, \citenamefont {Lee}, \citenamefont {Seok}, \citenamefont {Kim}, \citenamefont {Park}, \citenamefont {{\v{S}}mejkal} \emph {et~al.}}]{lee2024MnTe}%
  \BibitemOpen
  \bibfield  {author} {\bibinfo {author} {\bibfnamefont {S.}~\bibnamefont {Lee}}, \bibinfo {author} {\bibfnamefont {S.}~\bibnamefont {Lee}}, \bibinfo {author} {\bibfnamefont {S.}~\bibnamefont {Jung}}, \bibinfo {author} {\bibfnamefont {J.}~\bibnamefont {Jung}}, \bibinfo {author} {\bibfnamefont {D.}~\bibnamefont {Kim}}, \bibinfo {author} {\bibfnamefont {Y.}~\bibnamefont {Lee}}, \bibinfo {author} {\bibfnamefont {B.}~\bibnamefont {Seok}}, \bibinfo {author} {\bibfnamefont {J.}~\bibnamefont {Kim}}, \bibinfo {author} {\bibfnamefont {B.~G.}\ \bibnamefont {Park}}, \bibinfo {author} {\bibfnamefont {L.}~\bibnamefont {{\v{S}}mejkal}}, \emph {et~al.},\ }\bibfield  {title} {\bibinfo {title} {Broken kramers degeneracy in altermagnetic {MnTe}},\ }\href {https://journals.aps.org/prl/abstract/10.1103/PhysRevLett.132.036702} {\bibfield  {journal} {\bibinfo  {journal} {Phys. Rev. Lett.}\ }\textbf {\bibinfo {volume} {132}},\ \bibinfo {pages} {036702} (\bibinfo {year} {2024})}\BibitemShut {NoStop}%
\bibitem [{\citenamefont {Reimers}\ \emph {et~al.}(2024)\citenamefont {Reimers}, \citenamefont {Odenbreit}, \citenamefont {{\v{S}}mejkal}, \citenamefont {Strocov}, \citenamefont {Constantinou}, \citenamefont {Hellenes}, \citenamefont {Jaeschke~Ubiergo}, \citenamefont {Campos}, \citenamefont {Bharadwaj}, \citenamefont {Chakraborty} \emph {et~al.}}]{reimers2024CrSb}%
  \BibitemOpen
  \bibfield  {author} {\bibinfo {author} {\bibfnamefont {S.}~\bibnamefont {Reimers}}, \bibinfo {author} {\bibfnamefont {L.}~\bibnamefont {Odenbreit}}, \bibinfo {author} {\bibfnamefont {L.}~\bibnamefont {{\v{S}}mejkal}}, \bibinfo {author} {\bibfnamefont {V.~N.}\ \bibnamefont {Strocov}}, \bibinfo {author} {\bibfnamefont {P.}~\bibnamefont {Constantinou}}, \bibinfo {author} {\bibfnamefont {A.~B.}\ \bibnamefont {Hellenes}}, \bibinfo {author} {\bibfnamefont {R.}~\bibnamefont {Jaeschke~Ubiergo}}, \bibinfo {author} {\bibfnamefont {W.~H.}\ \bibnamefont {Campos}}, \bibinfo {author} {\bibfnamefont {V.~K.}\ \bibnamefont {Bharadwaj}}, \bibinfo {author} {\bibfnamefont {A.}~\bibnamefont {Chakraborty}}, \emph {et~al.},\ }\bibfield  {title} {\bibinfo {title} {Direct observation of altermagnetic band splitting in {CrSb} thin films},\ }\href {https://www.nature.com/articles/s41467-024-46476-5} {\bibfield  {journal} {\bibinfo  {journal} {Nat. Commun.}\ }\textbf {\bibinfo {volume} {15}},\ \bibinfo {pages} {2116} (\bibinfo {year}
  {2024})}\BibitemShut {NoStop}%
\bibitem [{\citenamefont {Zhou}\ \emph {et~al.}(2025)\citenamefont {Zhou}, \citenamefont {Cheng}, \citenamefont {Hu}, \citenamefont {Chu}, \citenamefont {Bai}, \citenamefont {Han}, \citenamefont {Liu}, \citenamefont {Pan},\ and\ \citenamefont {Song}}]{zhou2025manipulation}%
  \BibitemOpen
  \bibfield  {author} {\bibinfo {author} {\bibfnamefont {Z.}~\bibnamefont {Zhou}}, \bibinfo {author} {\bibfnamefont {X.}~\bibnamefont {Cheng}}, \bibinfo {author} {\bibfnamefont {M.}~\bibnamefont {Hu}}, \bibinfo {author} {\bibfnamefont {R.}~\bibnamefont {Chu}}, \bibinfo {author} {\bibfnamefont {H.}~\bibnamefont {Bai}}, \bibinfo {author} {\bibfnamefont {L.}~\bibnamefont {Han}}, \bibinfo {author} {\bibfnamefont {J.}~\bibnamefont {Liu}}, \bibinfo {author} {\bibfnamefont {F.}~\bibnamefont {Pan}},\ and\ \bibinfo {author} {\bibfnamefont {C.}~\bibnamefont {Song}},\ }\bibfield  {title} {\bibinfo {title} {Manipulation of the altermagnetic order in {CrSb} via crystal symmetry},\ }\href {https://doi.org/10.1038/s41586-024-08436-3} {\bibfield  {journal} {\bibinfo  {journal} {Nature}\ }\textbf {\bibinfo {volume} {638}},\ \bibinfo {pages} {645} (\bibinfo {year} {2025})}\BibitemShut {NoStop}%
\bibitem [{\citenamefont {Yang}\ \emph {et~al.}(2025{\natexlab{a}})\citenamefont {Yang}, \citenamefont {Li}, \citenamefont {Yang}, \citenamefont {Li}, \citenamefont {Zheng}, \citenamefont {Zhu}, \citenamefont {Pan}, \citenamefont {Xu}, \citenamefont {Cao}, \citenamefont {Zhao} \emph {et~al.}}]{yang2025three}%
  \BibitemOpen
  \bibfield  {author} {\bibinfo {author} {\bibfnamefont {G.}~\bibnamefont {Yang}}, \bibinfo {author} {\bibfnamefont {Z.}~\bibnamefont {Li}}, \bibinfo {author} {\bibfnamefont {S.}~\bibnamefont {Yang}}, \bibinfo {author} {\bibfnamefont {J.}~\bibnamefont {Li}}, \bibinfo {author} {\bibfnamefont {H.}~\bibnamefont {Zheng}}, \bibinfo {author} {\bibfnamefont {W.}~\bibnamefont {Zhu}}, \bibinfo {author} {\bibfnamefont {Z.}~\bibnamefont {Pan}}, \bibinfo {author} {\bibfnamefont {Y.}~\bibnamefont {Xu}}, \bibinfo {author} {\bibfnamefont {S.}~\bibnamefont {Cao}}, \bibinfo {author} {\bibfnamefont {W.}~\bibnamefont {Zhao}}, \emph {et~al.},\ }\bibfield  {title} {\bibinfo {title} {Three-dimensional mapping of the altermagnetic spin splitting in {CrSb}},\ }\href {https://doi.org/10.1038/s41467-025-56647-7} {\bibfield  {journal} {\bibinfo  {journal} {Nat. Commun.}\ }\textbf {\bibinfo {volume} {16}},\ \bibinfo {pages} {1442} (\bibinfo {year} {2025}{\natexlab{a}})}\BibitemShut {NoStop}%
\bibitem [{\citenamefont {Jiang}\ \emph {et~al.}(2025)\citenamefont {Jiang}, \citenamefont {Hu}, \citenamefont {Bai}, \citenamefont {Song}, \citenamefont {Mu}, \citenamefont {Qu}, \citenamefont {Li}, \citenamefont {Zhu}, \citenamefont {Pi}, \citenamefont {Wei} \emph {et~al.}}]{jiang2025metallic}%
  \BibitemOpen
  \bibfield  {author} {\bibinfo {author} {\bibfnamefont {B.}~\bibnamefont {Jiang}}, \bibinfo {author} {\bibfnamefont {M.}~\bibnamefont {Hu}}, \bibinfo {author} {\bibfnamefont {J.}~\bibnamefont {Bai}}, \bibinfo {author} {\bibfnamefont {Z.}~\bibnamefont {Song}}, \bibinfo {author} {\bibfnamefont {C.}~\bibnamefont {Mu}}, \bibinfo {author} {\bibfnamefont {G.}~\bibnamefont {Qu}}, \bibinfo {author} {\bibfnamefont {W.}~\bibnamefont {Li}}, \bibinfo {author} {\bibfnamefont {W.}~\bibnamefont {Zhu}}, \bibinfo {author} {\bibfnamefont {H.}~\bibnamefont {Pi}}, \bibinfo {author} {\bibfnamefont {Z.}~\bibnamefont {Wei}}, \emph {et~al.},\ }\bibfield  {title} {\bibinfo {title} {A metallic room-temperature d-wave altermagnet},\ }\href {https://doi.org/10.1038/s41567-025-02822-y} {\bibfield  {journal} {\bibinfo  {journal} {Nat. Phys.}\ }\textbf {\bibinfo {volume} {21}},\ \bibinfo {pages} {754} (\bibinfo {year} {2025})}\BibitemShut {NoStop}%
\bibitem [{\citenamefont {Zhang}\ \emph {et~al.}(2025)\citenamefont {Zhang}, \citenamefont {Cheng}, \citenamefont {Yin}, \citenamefont {Liu}, \citenamefont {Deng}, \citenamefont {Qiao}, \citenamefont {Shi}, \citenamefont {Zhang}, \citenamefont {Lin}, \citenamefont {Liu} \emph {et~al.}}]{zhang2025crystal}%
  \BibitemOpen
  \bibfield  {author} {\bibinfo {author} {\bibfnamefont {F.}~\bibnamefont {Zhang}}, \bibinfo {author} {\bibfnamefont {X.}~\bibnamefont {Cheng}}, \bibinfo {author} {\bibfnamefont {Z.}~\bibnamefont {Yin}}, \bibinfo {author} {\bibfnamefont {C.}~\bibnamefont {Liu}}, \bibinfo {author} {\bibfnamefont {L.}~\bibnamefont {Deng}}, \bibinfo {author} {\bibfnamefont {Y.}~\bibnamefont {Qiao}}, \bibinfo {author} {\bibfnamefont {Z.}~\bibnamefont {Shi}}, \bibinfo {author} {\bibfnamefont {S.}~\bibnamefont {Zhang}}, \bibinfo {author} {\bibfnamefont {J.}~\bibnamefont {Lin}}, \bibinfo {author} {\bibfnamefont {Z.}~\bibnamefont {Liu}}, \emph {et~al.},\ }\bibfield  {title} {\bibinfo {title} {Crystal-symmetry-paired spin--valley locking in a layered room-temperature metallic altermagnet candidate},\ }\href {https://doi.org/10.1038/s41567-025-02864-2} {\bibfield  {journal} {\bibinfo  {journal} {Nat. Phys.}\ }\textbf {\bibinfo {volume} {21}},\ \bibinfo {pages} {760} (\bibinfo {year} {2025})}\BibitemShut {NoStop}%
\bibitem [{\citenamefont {Hu}\ \emph {et~al.}(2025)\citenamefont {Hu}, \citenamefont {Song}, \citenamefont {Cheng}, \citenamefont {Qu}, \citenamefont {Li}, \citenamefont {Huang}, \citenamefont {Zhu}, \citenamefont {Zhang}, \citenamefont {Tian}, \citenamefont {Chen} \emph {et~al.}}]{hu2025pronounced}%
  \BibitemOpen
  \bibfield  {author} {\bibinfo {author} {\bibfnamefont {M.}~\bibnamefont {Hu}}, \bibinfo {author} {\bibfnamefont {Z.}~\bibnamefont {Song}}, \bibinfo {author} {\bibfnamefont {J.}~\bibnamefont {Cheng}}, \bibinfo {author} {\bibfnamefont {G.}~\bibnamefont {Qu}}, \bibinfo {author} {\bibfnamefont {Z.}~\bibnamefont {Li}}, \bibinfo {author} {\bibfnamefont {Y.}~\bibnamefont {Huang}}, \bibinfo {author} {\bibfnamefont {J.}~\bibnamefont {Zhu}}, \bibinfo {author} {\bibfnamefont {G.}~\bibnamefont {Zhang}}, \bibinfo {author} {\bibfnamefont {D.}~\bibnamefont {Tian}}, \bibinfo {author} {\bibfnamefont {L.}~\bibnamefont {Chen}}, \emph {et~al.},\ }\bibfield  {title} {\bibinfo {title} {Pronounced orbital-selective electron-electron correlation and electron-phonon coupling in {V$_2$Se$_2$O}},\ }\href {https://doi.org/10.48550/arXiv.2510.04657} {\bibfield  {journal} {\bibinfo  {journal} {arXiv preprint arXiv:2510.04657}\ } (\bibinfo {year} {2025})}\BibitemShut {NoStop}%
\bibitem [{\citenamefont {Chen}\ \emph {et~al.}(2025)\citenamefont {Chen}, \citenamefont {Yue}, \citenamefont {Cheng}, \citenamefont {Bai}, \citenamefont {Zhang}, \citenamefont {Ma}, \citenamefont {Hong}, \citenamefont {Chen}, \citenamefont {Wang}, \citenamefont {Wang} \emph {et~al.}}]{chen2025compression}%
  \BibitemOpen
  \bibfield  {author} {\bibinfo {author} {\bibfnamefont {L.}~\bibnamefont {Chen}}, \bibinfo {author} {\bibfnamefont {J.}~\bibnamefont {Yue}}, \bibinfo {author} {\bibfnamefont {J.}~\bibnamefont {Cheng}}, \bibinfo {author} {\bibfnamefont {J.}~\bibnamefont {Bai}}, \bibinfo {author} {\bibfnamefont {Z.}~\bibnamefont {Zhang}}, \bibinfo {author} {\bibfnamefont {X.}~\bibnamefont {Ma}}, \bibinfo {author} {\bibfnamefont {F.}~\bibnamefont {Hong}}, \bibinfo {author} {\bibfnamefont {G.}~\bibnamefont {Chen}}, \bibinfo {author} {\bibfnamefont {J.-T.}\ \bibnamefont {Wang}}, \bibinfo {author} {\bibfnamefont {Z.}~\bibnamefont {Wang}}, \emph {et~al.},\ }\bibfield  {title} {\bibinfo {title} {Compression-induced magnetic obstructed atomic insulator and spin singlet state in antiferromagnetic {KV$_2$Se$_2$O}},\ }\href {https://doi.org/10.48550/arXiv.2511.06712} {\bibfield  {journal} {\bibinfo  {journal} {arXiv preprint arXiv:2511.06712}\ } (\bibinfo {year} {2025})}\BibitemShut {NoStop}%
\bibitem [{\citenamefont {Liu}\ \emph {et~al.}(2025{\natexlab{d}})\citenamefont {Liu}, \citenamefont {Li}, \citenamefont {Liu}, \citenamefont {Lu}, \citenamefont {Li}, \citenamefont {Liu},\ and\ \citenamefont {Cao}}]{Liu2025prbPhysical}%
  \BibitemOpen
  \bibfield  {author} {\bibinfo {author} {\bibfnamefont {C.-C.}\ \bibnamefont {Liu}}, \bibinfo {author} {\bibfnamefont {J.}~\bibnamefont {Li}}, \bibinfo {author} {\bibfnamefont {J.-Y.}\ \bibnamefont {Liu}}, \bibinfo {author} {\bibfnamefont {J.-Y.}\ \bibnamefont {Lu}}, \bibinfo {author} {\bibfnamefont {H.-X.}\ \bibnamefont {Li}}, \bibinfo {author} {\bibfnamefont {Y.}~\bibnamefont {Liu}},\ and\ \bibinfo {author} {\bibfnamefont {G.-H.}\ \bibnamefont {Cao}},\ }\bibfield  {title} {\bibinfo {title} {Physical properties and first-principles calculations of an altermagnet candidate {Cs$_{1-\delta}$V$_2$Te$_2$O}},\ }\href {https://doi.org/10.1103/vch1-4khc} {\bibfield  {journal} {\bibinfo  {journal} {Phys. Rev. B}\ }\textbf {\bibinfo {volume} {112}},\ \bibinfo {pages} {224439} (\bibinfo {year} {2025}{\natexlab{d}})}\BibitemShut {NoStop}%
\bibitem [{\citenamefont {Sun}\ \emph {et~al.}(2025)\citenamefont {Sun}, \citenamefont {Huang}, \citenamefont {Cheng}, \citenamefont {Zhang}, \citenamefont {Li}, \citenamefont {Luo}, \citenamefont {Ma}, \citenamefont {Yang}, \citenamefont {Yang}, \citenamefont {Chen}, \citenamefont {Sun}, \citenamefont {Gutmann}, \citenamefont {Capelli}, \citenamefont {Shen}, \citenamefont {Hao}, \citenamefont {He}, \citenamefont {Chen},\ and\ \citenamefont {Li}}]{Sun2025antiferrom}%
  \BibitemOpen
  \bibfield  {author} {\bibinfo {author} {\bibfnamefont {Y.}~\bibnamefont {Sun}}, \bibinfo {author} {\bibfnamefont {Y.}~\bibnamefont {Huang}}, \bibinfo {author} {\bibfnamefont {J.}~\bibnamefont {Cheng}}, \bibinfo {author} {\bibfnamefont {S.}~\bibnamefont {Zhang}}, \bibinfo {author} {\bibfnamefont {Z.}~\bibnamefont {Li}}, \bibinfo {author} {\bibfnamefont {H.}~\bibnamefont {Luo}}, \bibinfo {author} {\bibfnamefont {X.}~\bibnamefont {Ma}}, \bibinfo {author} {\bibfnamefont {W.}~\bibnamefont {Yang}}, \bibinfo {author} {\bibfnamefont {J.}~\bibnamefont {Yang}}, \bibinfo {author} {\bibfnamefont {D.}~\bibnamefont {Chen}}, \bibinfo {author} {\bibfnamefont {K.}~\bibnamefont {Sun}}, \bibinfo {author} {\bibfnamefont {M.}~\bibnamefont {Gutmann}}, \bibinfo {author} {\bibfnamefont {S.~C.}\ \bibnamefont {Capelli}}, \bibinfo {author} {\bibfnamefont {F.}~\bibnamefont {Shen}}, \bibinfo {author} {\bibfnamefont {J.}~\bibnamefont {Hao}}, \bibinfo {author} {\bibfnamefont {L.}~\bibnamefont {He}}, \bibinfo {author} {\bibfnamefont
  {G.}~\bibnamefont {Chen}},\ and\ \bibinfo {author} {\bibfnamefont {S.}~\bibnamefont {Li}},\ }\bibfield  {title} {\bibinfo {title} {Antiferromagnetic structure of {KV$_2$Se$_2$O}: A neutron diffraction study},\ }\href {https://doi.org/10.1103/27n8-5q4l} {\bibfield  {journal} {\bibinfo  {journal} {Phys. Rev. B}\ }\textbf {\bibinfo {volume} {112}},\ \bibinfo {pages} {184416} (\bibinfo {year} {2025})}\BibitemShut {NoStop}%
\bibitem [{\citenamefont {Yang}\ \emph {et~al.}(2025{\natexlab{b}})\citenamefont {Yang}, \citenamefont {Chen}, \citenamefont {Liu}, \citenamefont {Li}, \citenamefont {Pan}, \citenamefont {Deng}, \citenamefont {Zheng}, \citenamefont {Tang}, \citenamefont {Zheng}, \citenamefont {Zhu} \emph {et~al.}}]{yang2025observation}%
  \BibitemOpen
  \bibfield  {author} {\bibinfo {author} {\bibfnamefont {G.}~\bibnamefont {Yang}}, \bibinfo {author} {\bibfnamefont {R.}~\bibnamefont {Chen}}, \bibinfo {author} {\bibfnamefont {C.}~\bibnamefont {Liu}}, \bibinfo {author} {\bibfnamefont {J.}~\bibnamefont {Li}}, \bibinfo {author} {\bibfnamefont {Z.}~\bibnamefont {Pan}}, \bibinfo {author} {\bibfnamefont {L.}~\bibnamefont {Deng}}, \bibinfo {author} {\bibfnamefont {N.}~\bibnamefont {Zheng}}, \bibinfo {author} {\bibfnamefont {Y.}~\bibnamefont {Tang}}, \bibinfo {author} {\bibfnamefont {H.}~\bibnamefont {Zheng}}, \bibinfo {author} {\bibfnamefont {W.}~\bibnamefont {Zhu}}, \emph {et~al.},\ }\bibfield  {title} {\bibinfo {title} {Observation of hidden altermagnetism in {Cs$_{1-\delta}$V$_2$Te$_2$O}},\ }\href {https://arxiv.org/abs/2512.00972} {\bibfield  {journal} {\bibinfo  {journal} {arXiv preprint arXiv:2512.00972}\ } (\bibinfo {year} {2025}{\natexlab{b}})}\BibitemShut {NoStop}%
\bibitem [{\citenamefont {Hu}\ \emph {et~al.}(2026)\citenamefont {Hu}, \citenamefont {Cheng}, \citenamefont {Duan}, \citenamefont {Hu}, \citenamefont {Jiang}, \citenamefont {Xiao}, \citenamefont {Li}, \citenamefont {Pan}, \citenamefont {Deng}, \citenamefont {Liu} \emph {et~al.}}]{hu2026observation}%
  \BibitemOpen
  \bibfield  {author} {\bibinfo {author} {\bibfnamefont {Q.}~\bibnamefont {Hu}}, \bibinfo {author} {\bibfnamefont {X.}~\bibnamefont {Cheng}}, \bibinfo {author} {\bibfnamefont {Q.}~\bibnamefont {Duan}}, \bibinfo {author} {\bibfnamefont {Y.}~\bibnamefont {Hu}}, \bibinfo {author} {\bibfnamefont {B.}~\bibnamefont {Jiang}}, \bibinfo {author} {\bibfnamefont {Y.}~\bibnamefont {Xiao}}, \bibinfo {author} {\bibfnamefont {Y.}~\bibnamefont {Li}}, \bibinfo {author} {\bibfnamefont {M.}~\bibnamefont {Pan}}, \bibinfo {author} {\bibfnamefont {L.}~\bibnamefont {Deng}}, \bibinfo {author} {\bibfnamefont {C.}~\bibnamefont {Liu}}, \emph {et~al.},\ }\bibfield  {title} {\bibinfo {title} {Observation of spin-valley locked nodal lines in a {quasi-2D} altermagnet},\ }\href {https://doi.org/10.48550/arXiv.2601.02883} {\bibfield  {journal} {\bibinfo  {journal} {arXiv preprint arXiv:2601.02883}\ } (\bibinfo {year} {2026})}\BibitemShut {NoStop}%
\bibitem [{\citenamefont {Ma}\ \emph {et~al.}(2021)\citenamefont {Ma}, \citenamefont {Hu}, \citenamefont {Li}, \citenamefont {Liu}, \citenamefont {Yao}, \citenamefont {Jia},\ and\ \citenamefont {Liu}}]{ma2021multifunctional}%
  \BibitemOpen
  \bibfield  {author} {\bibinfo {author} {\bibfnamefont {H.-Y.}\ \bibnamefont {Ma}}, \bibinfo {author} {\bibfnamefont {M.}~\bibnamefont {Hu}}, \bibinfo {author} {\bibfnamefont {N.}~\bibnamefont {Li}}, \bibinfo {author} {\bibfnamefont {J.}~\bibnamefont {Liu}}, \bibinfo {author} {\bibfnamefont {W.}~\bibnamefont {Yao}}, \bibinfo {author} {\bibfnamefont {J.-F.}\ \bibnamefont {Jia}},\ and\ \bibinfo {author} {\bibfnamefont {J.}~\bibnamefont {Liu}},\ }\bibfield  {title} {\bibinfo {title} {Multifunctional antiferromagnetic materials with giant piezomagnetism and noncollinear spin current},\ }\href {https://doi.org/10.1038/s41467-021-23127-7} {\bibfield  {journal} {\bibinfo  {journal} {Nat. Commun.}\ }\textbf {\bibinfo {volume} {12}},\ \bibinfo {pages} {2846} (\bibinfo {year} {2021})}\BibitemShut {NoStop}%
\bibitem [{\citenamefont {Lai}\ \emph {et~al.}(2025)\citenamefont {Lai}, \citenamefont {Yu}, \citenamefont {Liu}, \citenamefont {Liu}, \citenamefont {Xing}, \citenamefont {Chen},\ and\ \citenamefont {Sun}}]{Lai2025dwave}%
  \BibitemOpen
  \bibfield  {author} {\bibinfo {author} {\bibfnamefont {J.}~\bibnamefont {Lai}}, \bibinfo {author} {\bibfnamefont {T.}~\bibnamefont {Yu}}, \bibinfo {author} {\bibfnamefont {P.}~\bibnamefont {Liu}}, \bibinfo {author} {\bibfnamefont {L.}~\bibnamefont {Liu}}, \bibinfo {author} {\bibfnamefont {G.}~\bibnamefont {Xing}}, \bibinfo {author} {\bibfnamefont {X.-Q.}\ \bibnamefont {Chen}},\ and\ \bibinfo {author} {\bibfnamefont {Y.}~\bibnamefont {Sun}},\ }\bibfield  {title} {\bibinfo {title} {$d$-wave flat {Fermi} surface in altermagnets enables maximum charge-to-spin conversion},\ }\href {https://doi.org/10.1103/bf1n-sxdl} {\bibfield  {journal} {\bibinfo  {journal} {Phys. Rev. Lett.}\ }\textbf {\bibinfo {volume} {135}},\ \bibinfo {pages} {256702} (\bibinfo {year} {2025})}\BibitemShut {NoStop}%
\bibitem [{\citenamefont {Li}\ \emph {et~al.}(2026)\citenamefont {Li}, \citenamefont {Hou}, \citenamefont {Zhu}, \citenamefont {Zheng}, \citenamefont {Song}, \citenamefont {Liu}, \citenamefont {Zhang},\ and\ \citenamefont {Hu}}]{li2026JJevenodd}%
  \BibitemOpen
  \bibfield  {author} {\bibinfo {author} {\bibfnamefont {C.}~\bibnamefont {Li}}, \bibinfo {author} {\bibfnamefont {J.-X.}\ \bibnamefont {Hou}}, \bibinfo {author} {\bibfnamefont {S.-L.}\ \bibnamefont {Zhu}}, \bibinfo {author} {\bibfnamefont {H.}~\bibnamefont {Zheng}}, \bibinfo {author} {\bibfnamefont {Y.}~\bibnamefont {Song}}, \bibinfo {author} {\bibfnamefont {Y.}~\bibnamefont {Liu}}, \bibinfo {author} {\bibfnamefont {S.-B.}\ \bibnamefont {Zhang}},\ and\ \bibinfo {author} {\bibfnamefont {L.-H.}\ \bibnamefont {Hu}},\ }\bibfield  {title} {\bibinfo {title} {Altermagnetic even-odd effects in {CsV$_2$Te$_2$O} {Josephson} junctions},\ }\href {https://arxiv.org/abs/2602.14485} {\bibfield  {journal} {\bibinfo  {journal} {arXiv preprint arXiv:2602.14485}\ } (\bibinfo {year} {2026})}\BibitemShut {NoStop}%
\bibitem [{\citenamefont {Zhang}\ \emph {et~al.}(2022)\citenamefont {Zhang}, \citenamefont {Pincelli}, \citenamefont {Jozwiak}, \citenamefont {Kondo}, \citenamefont {Ernstorfer}, \citenamefont {Sato},\ and\ \citenamefont {Zhou}}]{zhang2022Angleresolved}%
  \BibitemOpen
  \bibfield  {author} {\bibinfo {author} {\bibfnamefont {H.}~\bibnamefont {Zhang}}, \bibinfo {author} {\bibfnamefont {T.}~\bibnamefont {Pincelli}}, \bibinfo {author} {\bibfnamefont {C.}~\bibnamefont {Jozwiak}}, \bibinfo {author} {\bibfnamefont {T.}~\bibnamefont {Kondo}}, \bibinfo {author} {\bibfnamefont {R.}~\bibnamefont {Ernstorfer}}, \bibinfo {author} {\bibfnamefont {T.}~\bibnamefont {Sato}},\ and\ \bibinfo {author} {\bibfnamefont {S.}~\bibnamefont {Zhou}},\ }\bibfield  {title} {\bibinfo {title} {Angle-resolved photoemission spectroscopy},\ }\href {https://doi.org/10.1038/s43586-022-00133-7} {\bibfield  {journal} {\bibinfo  {journal} {Nat. Rev. Methods Prim.}\ }\textbf {\bibinfo {volume} {2}},\ \bibinfo {pages} {54} (\bibinfo {year} {2022})}\BibitemShut {NoStop}%
\bibitem [{\citenamefont {Liu}\ \emph {et~al.}(2026)\citenamefont {Liu}, \citenamefont {Ma}, \citenamefont {Zhang}, \citenamefont {Jing}, \citenamefont {Liu},\ and\ \citenamefont {Shen}}]{liuSymmetry2026}%
  \BibitemOpen
  \bibfield  {author} {\bibinfo {author} {\bibfnamefont {J.}~\bibnamefont {Liu}}, \bibinfo {author} {\bibfnamefont {X.}~\bibnamefont {Ma}}, \bibinfo {author} {\bibfnamefont {X.}~\bibnamefont {Zhang}}, \bibinfo {author} {\bibfnamefont {W.}~\bibnamefont {Jing}}, \bibinfo {author} {\bibfnamefont {Z.}~\bibnamefont {Liu}},\ and\ \bibinfo {author} {\bibfnamefont {D.}~\bibnamefont {Shen}},\ }\bibfield  {title} {\bibinfo {title} {Symmetry-driven spin splitting in altermagnets: An angle-resolved photoemission spectroscopy perspective},\ }\href {https://doi.org/10.1186/s40580-026-00536-2} {\bibfield  {journal} {\bibinfo  {journal} {Nano Converg.}\ }\textbf {\bibinfo {volume} {13}},\ \bibinfo {pages} {6} (\bibinfo {year} {2026})}\BibitemShut {NoStop}%
\bibitem [{\citenamefont {Bode}(2003)}]{bodeSpin2003}%
  \BibitemOpen
  \bibfield  {author} {\bibinfo {author} {\bibfnamefont {M.}~\bibnamefont {Bode}},\ }\bibfield  {title} {\bibinfo {title} {Spin-polarized scanning tunnelling microscopy},\ }\href {https://doi.org/10.1088/0034-4885/66/4/203} {\bibfield  {journal} {\bibinfo  {journal} {Rep. Prog. Phys.}\ }\textbf {\bibinfo {volume} {66}},\ \bibinfo {pages} {523} (\bibinfo {year} {2003})}\BibitemShut {NoStop}%
\bibitem [{\citenamefont {Wiesendanger}(2009)}]{wiesendangerSpinMappingNanoscale2009}%
  \BibitemOpen
  \bibfield  {author} {\bibinfo {author} {\bibfnamefont {R.}~\bibnamefont {Wiesendanger}},\ }\bibfield  {title} {\bibinfo {title} {Spin mapping at the nanoscale and atomic scale},\ }\href {https://doi.org/10.1103/RevModPhys.81.1495} {\bibfield  {journal} {\bibinfo  {journal} {Rev. Mod. Phys.}\ }\textbf {\bibinfo {volume} {81}},\ \bibinfo {pages} {1495} (\bibinfo {year} {2009})}\BibitemShut {NoStop}%
\bibitem [{\citenamefont {Dzero}\ \emph {et~al.}(2010)\citenamefont {Dzero}, \citenamefont {Sun}, \citenamefont {Galitski},\ and\ \citenamefont {Coleman}}]{Dzero2010Topo}%
  \BibitemOpen
  \bibfield  {author} {\bibinfo {author} {\bibfnamefont {M.}~\bibnamefont {Dzero}}, \bibinfo {author} {\bibfnamefont {K.}~\bibnamefont {Sun}}, \bibinfo {author} {\bibfnamefont {V.}~\bibnamefont {Galitski}},\ and\ \bibinfo {author} {\bibfnamefont {P.}~\bibnamefont {Coleman}},\ }\bibfield  {title} {\bibinfo {title} {Topological {Kondo} insulators},\ }\href {https://link.aps.org/doi/10.1103/PhysRevLett.104.106408} {\bibfield  {journal} {\bibinfo  {journal} {Phys. Rev. Lett.}\ }\textbf {\bibinfo {volume} {104}},\ \bibinfo {pages} {106408} (\bibinfo {year} {2010})}\BibitemShut {NoStop}%
\bibitem [{\citenamefont {Aishwarya}\ \emph {et~al.}(2022)\citenamefont {Aishwarya}, \citenamefont {Cai}, \citenamefont {Raghavan}, \citenamefont {Romanelli}, \citenamefont {Wang}, \citenamefont {Li}, \citenamefont {Gu}, \citenamefont {Hirsbrunner}, \citenamefont {Hughes}, \citenamefont {Liu} \emph {et~al.}}]{aishwarya2022spin}%
  \BibitemOpen
  \bibfield  {author} {\bibinfo {author} {\bibfnamefont {A.}~\bibnamefont {Aishwarya}}, \bibinfo {author} {\bibfnamefont {Z.}~\bibnamefont {Cai}}, \bibinfo {author} {\bibfnamefont {A.}~\bibnamefont {Raghavan}}, \bibinfo {author} {\bibfnamefont {M.}~\bibnamefont {Romanelli}}, \bibinfo {author} {\bibfnamefont {X.}~\bibnamefont {Wang}}, \bibinfo {author} {\bibfnamefont {X.}~\bibnamefont {Li}}, \bibinfo {author} {\bibfnamefont {G.}~\bibnamefont {Gu}}, \bibinfo {author} {\bibfnamefont {M.}~\bibnamefont {Hirsbrunner}}, \bibinfo {author} {\bibfnamefont {T.}~\bibnamefont {Hughes}}, \bibinfo {author} {\bibfnamefont {F.}~\bibnamefont {Liu}}, \emph {et~al.},\ }\bibfield  {title} {\bibinfo {title} {Spin-selective tunneling from nanowires of the candidate topological {Kondo} insulator {SmB$_6$}},\ }\href {http://dx.doi.org/10.1126/science.abj8765} {\bibfield  {journal} {\bibinfo  {journal} {Science}\ }\textbf {\bibinfo {volume} {377}},\ \bibinfo {pages} {1218} (\bibinfo {year} {2022})}\BibitemShut {NoStop}%
\bibitem [{\citenamefont {Banerjee}\ \emph {et~al.}(2025)\citenamefont {Banerjee}, \citenamefont {Aishwarya}, \citenamefont {Liu}, \citenamefont {Jiao}, \citenamefont {Madhavan}, \citenamefont {Mele},\ and\ \citenamefont {Coleman}}]{banerjee_axionic_2025}%
  \BibitemOpen
  \bibfield  {author} {\bibinfo {author} {\bibfnamefont {S.}~\bibnamefont {Banerjee}}, \bibinfo {author} {\bibfnamefont {A.}~\bibnamefont {Aishwarya}}, \bibinfo {author} {\bibfnamefont {F.}~\bibnamefont {Liu}}, \bibinfo {author} {\bibfnamefont {L.}~\bibnamefont {Jiao}}, \bibinfo {author} {\bibfnamefont {V.}~\bibnamefont {Madhavan}}, \bibinfo {author} {\bibfnamefont {E.~J.}\ \bibnamefont {Mele}},\ and\ \bibinfo {author} {\bibfnamefont {P.}~\bibnamefont {Coleman}},\ }\href {https://doi.org/10.48550/arXiv.2512.05057} {\bibinfo {title} {Axionic tunneling from a topological {{Kondo}} insulator}} (\bibinfo {year} {2025}),\ \Eprint {https://arxiv.org/abs/2512.05057} {arXiv:2512.05057 [cond-mat.mes-hall]} \BibitemShut {NoStop}%
\bibitem [{\citenamefont {Yan}\ \emph {et~al.}(2026)\citenamefont {Yan}, \citenamefont {Song}, \citenamefont {Song}, \citenamefont {Fang}, \citenamefont {Weng},\ and\ \citenamefont {Wu}}]{yan2026magnetic}%
  \BibitemOpen
  \bibfield  {author} {\bibinfo {author} {\bibfnamefont {X.}~\bibnamefont {Yan}}, \bibinfo {author} {\bibfnamefont {Z.}~\bibnamefont {Song}}, \bibinfo {author} {\bibfnamefont {J.}~\bibnamefont {Song}}, \bibinfo {author} {\bibfnamefont {Z.}~\bibnamefont {Fang}}, \bibinfo {author} {\bibfnamefont {H.}~\bibnamefont {Weng}},\ and\ \bibinfo {author} {\bibfnamefont {Q.}~\bibnamefont {Wu}},\ }\bibfield  {title} {\bibinfo {title} {Magnetic symmetry breaking driven ``inverse magnetic breakdown'' in a d-wave altermagnet {KV$_2$Se$_2$O}},\ }\href {https://doi.org/10.1007/s11433-025-2881-5} {\bibfield  {journal} {\bibinfo  {journal} {Sci. China Phys. Mech. Astron.}\ }\textbf {\bibinfo {volume} {69}},\ \bibinfo {pages} {257011} (\bibinfo {year} {2026})}\BibitemShut {NoStop}%
\bibitem [{\citenamefont {Bai}\ \emph {et~al.}(2024{\natexlab{b}})\citenamefont {Bai}, \citenamefont {Ruan}, \citenamefont {Dong}, \citenamefont {Zhang}, \citenamefont {Liu}, \citenamefont {Cheng}, \citenamefont {Liu}, \citenamefont {Li}, \citenamefont {Sun}, \citenamefont {Huang}, \citenamefont {Ren},\ and\ \citenamefont {Chen}}]{baiAbsence2024}%
  \BibitemOpen
  \bibfield  {author} {\bibinfo {author} {\bibfnamefont {J.}~\bibnamefont {Bai}}, \bibinfo {author} {\bibfnamefont {B.}~\bibnamefont {Ruan}}, \bibinfo {author} {\bibfnamefont {Q.}~\bibnamefont {Dong}}, \bibinfo {author} {\bibfnamefont {L.}~\bibnamefont {Zhang}}, \bibinfo {author} {\bibfnamefont {Q.}~\bibnamefont {Liu}}, \bibinfo {author} {\bibfnamefont {J.}~\bibnamefont {Cheng}}, \bibinfo {author} {\bibfnamefont {P.}~\bibnamefont {Liu}}, \bibinfo {author} {\bibfnamefont {C.}~\bibnamefont {Li}}, \bibinfo {author} {\bibfnamefont {Y.}~\bibnamefont {Sun}}, \bibinfo {author} {\bibfnamefont {Y.}~\bibnamefont {Huang}}, \bibinfo {author} {\bibfnamefont {Z.}~\bibnamefont {Ren}},\ and\ \bibinfo {author} {\bibfnamefont {G.}~\bibnamefont {Chen}},\ }\bibfield  {title} {\bibinfo {title} {Absence of long-range order in the vanadium oxychalcogenide {KV$_{2}$Se$_{2}$O} with nontrivial band topology},\ }\href {https://doi.org/10.1103/PhysRevB.110.165151} {\bibfield  {journal} {\bibinfo  {journal} {Phys. Rev. B}\ }\textbf
  {\bibinfo {volume} {110}},\ \bibinfo {pages} {165151} (\bibinfo {year} {2024}{\natexlab{b}})}\BibitemShut {NoStop}%
\bibitem [{\citenamefont {Zhuang}\ \emph {et~al.}(2025)\citenamefont {Zhuang}, \citenamefont {Bai}, \citenamefont {Cheng}, \citenamefont {Li}, \citenamefont {Meng}, \citenamefont {Wang}, \citenamefont {Zhang}, \citenamefont {Shen}, \citenamefont {Wang}, \citenamefont {Chen},\ and\ \citenamefont {Yu}}]{zhuangCharge2025a}%
  \BibitemOpen
  \bibfield  {author} {\bibinfo {author} {\bibfnamefont {H.}~\bibnamefont {Zhuang}}, \bibinfo {author} {\bibfnamefont {J.}~\bibnamefont {Bai}}, \bibinfo {author} {\bibfnamefont {J.}~\bibnamefont {Cheng}}, \bibinfo {author} {\bibfnamefont {X.}~\bibnamefont {Li}}, \bibinfo {author} {\bibfnamefont {Y.}~\bibnamefont {Meng}}, \bibinfo {author} {\bibfnamefont {L.}~\bibnamefont {Wang}}, \bibinfo {author} {\bibfnamefont {Q.}~\bibnamefont {Zhang}}, \bibinfo {author} {\bibfnamefont {X.}~\bibnamefont {Shen}}, \bibinfo {author} {\bibfnamefont {Y.}~\bibnamefont {Wang}}, \bibinfo {author} {\bibfnamefont {G.}~\bibnamefont {Chen}},\ and\ \bibinfo {author} {\bibfnamefont {R.}~\bibnamefont {Yu}},\ }\bibfield  {title} {\bibinfo {title} {Charge transfer caused anomalies of physical properties of {KV$_{2}$Se$_{2}$O}},\ }\href {https://doi.org/10.1209/0295-5075/adb44e} {\bibfield  {journal} {\bibinfo  {journal} {Europhys. Lett.}\ }\textbf {\bibinfo {volume} {150}},\ \bibinfo {pages} {36003} (\bibinfo {year} {2025})}\BibitemShut
  {NoStop}%
\bibitem [{\citenamefont {Xu}\ \emph {et~al.}(2025)\citenamefont {Xu}, \citenamefont {Zhang}, \citenamefont {Feng},\ and\ \citenamefont {Tian}}]{xu2025prb}%
  \BibitemOpen
  \bibfield  {author} {\bibinfo {author} {\bibfnamefont {Y.}~\bibnamefont {Xu}}, \bibinfo {author} {\bibfnamefont {H.}~\bibnamefont {Zhang}}, \bibinfo {author} {\bibfnamefont {M.}~\bibnamefont {Feng}},\ and\ \bibinfo {author} {\bibfnamefont {F.}~\bibnamefont {Tian}},\ }\bibfield  {title} {\bibinfo {title} {Electronic structure, magnetic transition, and {Fermi} surface instability of the room-temperature altermagnet {KV$_{2}$Se$_{2}$O}},\ }\href {https://doi.org/10.1103/r8nc-dpt8} {\bibfield  {journal} {\bibinfo  {journal} {Phys. Rev. B}\ }\textbf {\bibinfo {volume} {112}},\ \bibinfo {pages} {125141} (\bibinfo {year} {2025})}\BibitemShut {NoStop}%
\bibitem [{\citenamefont {Wang}\ \emph {et~al.}(2025)\citenamefont {Wang}, \citenamefont {Yu}, \citenamefont {Cheng}, \citenamefont {Xiao}, \citenamefont {Ma}, \citenamefont {Quan}, \citenamefont {Song}, \citenamefont {Zhang}, \citenamefont {Zhang}, \citenamefont {Ma}, \citenamefont {Liu}, \citenamefont {Yadav}, \citenamefont {Shi}, \citenamefont {Wang}, \citenamefont {Niu}, \citenamefont {Gao}, \citenamefont {Xiang}, \citenamefont {Liu}, \citenamefont {Wang},\ and\ \citenamefont {Chen}}]{wang2025atomic}%
  \BibitemOpen
  \bibfield  {author} {\bibinfo {author} {\bibfnamefont {Z.}~\bibnamefont {Wang}}, \bibinfo {author} {\bibfnamefont {S.}~\bibnamefont {Yu}}, \bibinfo {author} {\bibfnamefont {X.}~\bibnamefont {Cheng}}, \bibinfo {author} {\bibfnamefont {X.}~\bibnamefont {Xiao}}, \bibinfo {author} {\bibfnamefont {W.}~\bibnamefont {Ma}}, \bibinfo {author} {\bibfnamefont {F.}~\bibnamefont {Quan}}, \bibinfo {author} {\bibfnamefont {H.}~\bibnamefont {Song}}, \bibinfo {author} {\bibfnamefont {K.}~\bibnamefont {Zhang}}, \bibinfo {author} {\bibfnamefont {Y.}~\bibnamefont {Zhang}}, \bibinfo {author} {\bibfnamefont {Y.}~\bibnamefont {Ma}}, \bibinfo {author} {\bibfnamefont {W.}~\bibnamefont {Liu}}, \bibinfo {author} {\bibfnamefont {P.}~\bibnamefont {Yadav}}, \bibinfo {author} {\bibfnamefont {X.}~\bibnamefont {Shi}}, \bibinfo {author} {\bibfnamefont {Z.}~\bibnamefont {Wang}}, \bibinfo {author} {\bibfnamefont {Q.}~\bibnamefont {Niu}}, \bibinfo {author} {\bibfnamefont {Y.}~\bibnamefont {Gao}}, \bibinfo {author} {\bibfnamefont
  {B.}~\bibnamefont {Xiang}}, \bibinfo {author} {\bibfnamefont {J.}~\bibnamefont {Liu}}, \bibinfo {author} {\bibfnamefont {Z.}~\bibnamefont {Wang}},\ and\ \bibinfo {author} {\bibfnamefont {X.}~\bibnamefont {Chen}},\ }\bibfield  {title} {\bibinfo {title} {Atomic-scale spin sensing of a {2D} $d$-wave altermagnet via helical tunneling},\ }\href {https://arxiv.org/abs/2512.23290} {\bibfield  {journal} {\bibinfo  {journal} {arXiv preprint arXiv:2512.23290}\ } (\bibinfo {year} {2025})}\BibitemShut {NoStop}%
\bibitem [{\citenamefont {Fu}\ \emph {et~al.}(2025)\citenamefont {Fu}, \citenamefont {Yang}, \citenamefont {Xiao}, \citenamefont {Wang}, \citenamefont {Wang}, \citenamefont {Yao}, \citenamefont {Xue},\ and\ \citenamefont {Li}}]{fu2025atomic}%
  \BibitemOpen
  \bibfield  {author} {\bibinfo {author} {\bibfnamefont {D.}~\bibnamefont {Fu}}, \bibinfo {author} {\bibfnamefont {L.}~\bibnamefont {Yang}}, \bibinfo {author} {\bibfnamefont {K.}~\bibnamefont {Xiao}}, \bibinfo {author} {\bibfnamefont {Y.}~\bibnamefont {Wang}}, \bibinfo {author} {\bibfnamefont {Z.}~\bibnamefont {Wang}}, \bibinfo {author} {\bibfnamefont {Y.}~\bibnamefont {Yao}}, \bibinfo {author} {\bibfnamefont {Q.-K.}\ \bibnamefont {Xue}},\ and\ \bibinfo {author} {\bibfnamefont {W.}~\bibnamefont {Li}},\ }\bibfield  {title} {\bibinfo {title} {Atomic-scale visualization of d-wave altermagnetism},\ }\href {https://doi.org/10.48550/arXiv.2512.24114} {\bibfield  {journal} {\bibinfo  {journal} {arXiv preprint arXiv:2512.24114}\ } (\bibinfo {year} {2025})}\BibitemShut {NoStop}%
\end{thebibliography}

%

\end{document}